\documentclass[12pt,preprint]{aastex}
\shorttitle{Wavelet Analyses of AGN}
\shortauthors{Espaillat et al.}

\begin{document}

\title{Wavelet Analysis of AGN X-ray Time Series: A QPO in 3C 273?}
\author{C. Espaillat\altaffilmark{1}, J. Bregman\altaffilmark{1}, P. Hughes\altaffilmark{1}, and E. Lloyd-Davies\altaffilmark{2}}

\altaffiltext{1}{Department of Astronomy, University of Michigan, 830 Dennison Building, 500 Church Street, Ann Arbor, MI 48109, USA; ccespa@umich.edu; jbregman@umich.edu; phughes@umich.edu}
\altaffiltext{2}{Astronomy Centre, University of Sussex, Falmer, Brighton, BN1 9QH, UK; E.Lloyd-Davies@sussex.ac.uk}

\begin{abstract}

  Quasi-periodic signals have yielded important constraints on the
  masses of black holes in galactic X-ray binaries, and here we extend
  this to active galactic nuclei (AGN).  We employ a wavelet technique
  to analyze 19 observations of 10 AGN obtained with the
  {\it XMM-Newton} EPIC-PN camera. We report the detection of a
  candidate 3.3 kilosecond quasi-period in
  3C 273.  If this period represents an orbital timescale originating
  near a last stable orbit of 3 $R_S$, it implies a central black hole
  mass of $7.3\times 10^6$ M$_\sun$.  For a maximally rotating black
  hole with a last stable orbit of 0.6 $R_S$, a central black hole
  mass of $8.1\times 10^7$ M$_\sun$ is implied.  Both of these estimates
  are substantially lower than previous reverberation mapping results
  which place the central black hole mass of 3C 273 at about $2.35\times 10^8$
  M$_\sun$.  Assuming that this reverberation mass is correct, the
  X-ray quasi-period would be caused by a higher order oscillatory mode
  of the accretion disk.

\end{abstract}

\keywords{galaxies: individual(\objectname{3C 273}) --- galaxies: active --- X-rays: galaxies}

\section{Introduction}

Quasi-periodic oscillations (QPOs) are thought to originate in the inner accretion disk of a black hole or neutron star in an X-ray binary (XRB) system \citep{vdk00}.  Consequently, QPOs have been used in galactic XRBs to introduce important constraints on the masses of the central black holes of these systems.

Previous work has revealed that AGN and XRBs are alike: noise power spectra have shown that similar physical processes may be underlying the X-ray variability in both \citep{ede99,utt02, mar03, vau03, mchar04, mchar05}.  Taking this resemblance into account and assuming that accretion onto a stellar-mass black hole is comparable to accretion onto a supermassive black hole, one would expect some AGN to exhibit QPOs similar to those observed in XRBs.  In supermassive black holes ($10^6$-$10^9$ M$_\sun$), these QPOs would be at much lower frequencies than those we find in stellar-mass black holes ($\sim$ 10 M$_\sun$).  Low frequency quasi-periods (LF QPOs) in XRBs range from 50 mHz to 30 Hz; scaling from a $\sim 1$ Hz QPO in a 10 $M_\sun$ XRB, a LF QPO in an AGN would occur on timescales of days to months \citep{vau05}, too long to be detectable for the AGN in our sample.  On the other hand, high frequency QPOs (HF QPOs) in XRBs have values of $\geq$ 100 Hz and assuming a 1/$M_{BH}$ scaling of frequencies, $f_{HFQPO}\sim 3\times 10^{-3} (M_{BH}/ 10^{6} M_\sun)^{-1}$ Hz \citep{abram}, corresponding to timescales greater than 400s for AGN.

While this parallel between AGN and XRBs seems promising, no claim of an X-ray quasi-period in an AGN has been found to be statistically robust.  \citet{vaub} remark that a major source of false detections arise from assuming an inappropriate background noise power spectrum.  X-ray variations of AGN have intrinsically red noise power spectra (i.e. the power spectra have a continuum resembling a power law with a steep slope; Press 1978), however many purported QPOs in AGN are compared against an assumed background of white noise (i.e. Poisson photon noise or a flat spectrum).  For example, in a $\sim$5 day ASCA observation of IRAS 18325-5926 the significance of the candidate periodicity was estimated with white noise \citep{iwa98}.  After including red noise in the periodogram fitting, \citet{vau} found that the candidate periodicity was no longer significant at the 95$\%$ level.   \citet{fiore} also claimed high ($>99\%$) significance peaks in NGC 4151, however, after fitting red noise and Poisson photon noise components of the spectrum \citet{vaub} showed that the significances of the QPOs fell below the 95$\%$ confidence level.  It is also difficult to constrain the significance of possible QPOs due to power spectra effects \citep{vaub}.  EXOSAT data of NGC 5548 were reported to have a significant period \citep{pap93}, but \citet{tag96} later showed that the significance of the candidate QPO was lower than previously reported once the uncertainties in modeling the spectrum were taken into consideration. 

This lack of statistically significant evidence for QPOs in AGN has led to questions of whether existing X-ray observations of AGN are sensitive enough to detect QPOs even if they are present \citep{vau05}.

Here we use a different technique to search for significant periodic
structures in the time variability data that have been collected for
AGNs with XMM-Newton. We use a wavelet transform technique, which can
have certain advantages relative to periodograms and Fourier power
spectra, the methods that have previously dominated the literature. The
wavelet technique, which has become widely used in other branches of
science, is particularly useful in identifying signals where the period
or its amplitude changes with time. This technique is applied to the
XMM-Newton data from bright 10 AGNs, with special care taken to properly
treat the noise characteristics and error analysis, and we find
a candidate 3.3 ks quasi-period in 3C 273.  

In Section 2, we present our observations and data reduction steps.  In Section 3, we provide an overview of the two wavelet techniques used in our analysis: the continuous wavelet transform and the cross-wavelet transform.  The results of these two techniques as well as significance tests are presented.  We also discuss structure function analysis for the AGN in our sample.  In Section 4 we argue that this 3.3 ks quasi-period in 3C 273 is consistent with what we would expect from oscillations in the accretion disk around the supermassive black hole based on current black hole mass estimates.

\section{Observations and Data Reduction} \label{obssection}

The 10 AGN in our sample were selected because they are bright and have {\it XMM-Newton} EPIC-PN camera observations which exceed 30 kiloseconds (ks).  In total, we have 19 observations and each observation's ID, date, length, and average counts are listed in Table~\ref{obslog}.  All observations are in the energy range 0.75 to 10 keV and most were taken in small window mode, which has a readout time of 6 milliseconds (ms).  The only exception is NGC 4151 Observation ID (Obs. ID): 0112830201, which was taken in full frame mode with a readout time of 73.4 ms.

Observation Data Files (ODFs) were obtained from the on-line {\it XMM-Newton} Science Archive and later reduced with the {\it XMM-Newton} Science Analysis Software (SAS, v. 7.0.0, 6.1.0, 5.4.1).  Source light curves, with 5 s bins, were extracted for a circular region centered on the source ($\sim$20$^\prime$$^\prime$).  Background light curves were obtained from a nearby rectangular source-free region and subtracted from the source light curves.  These rectangular background regions were larger than the source regions and were accordingly scaled down.  Due to strong flaring, the last few kiloseconds of data are excluded from most observations.  The count rates for the target sources are orders of magnitude greater
than the background count rate in the detection cell, so a rise in the
background is unimportant. We removed these last few kiloseconds of data
from the data stream just to be very cautious.  We note that in the observation of 3C 273 with the claimed detection, including the periods with flaring does not change our results.

\begin{figure}
\figurenum{1}
%\plotone{mkn421_epat_rescaled.ps}
\plotone{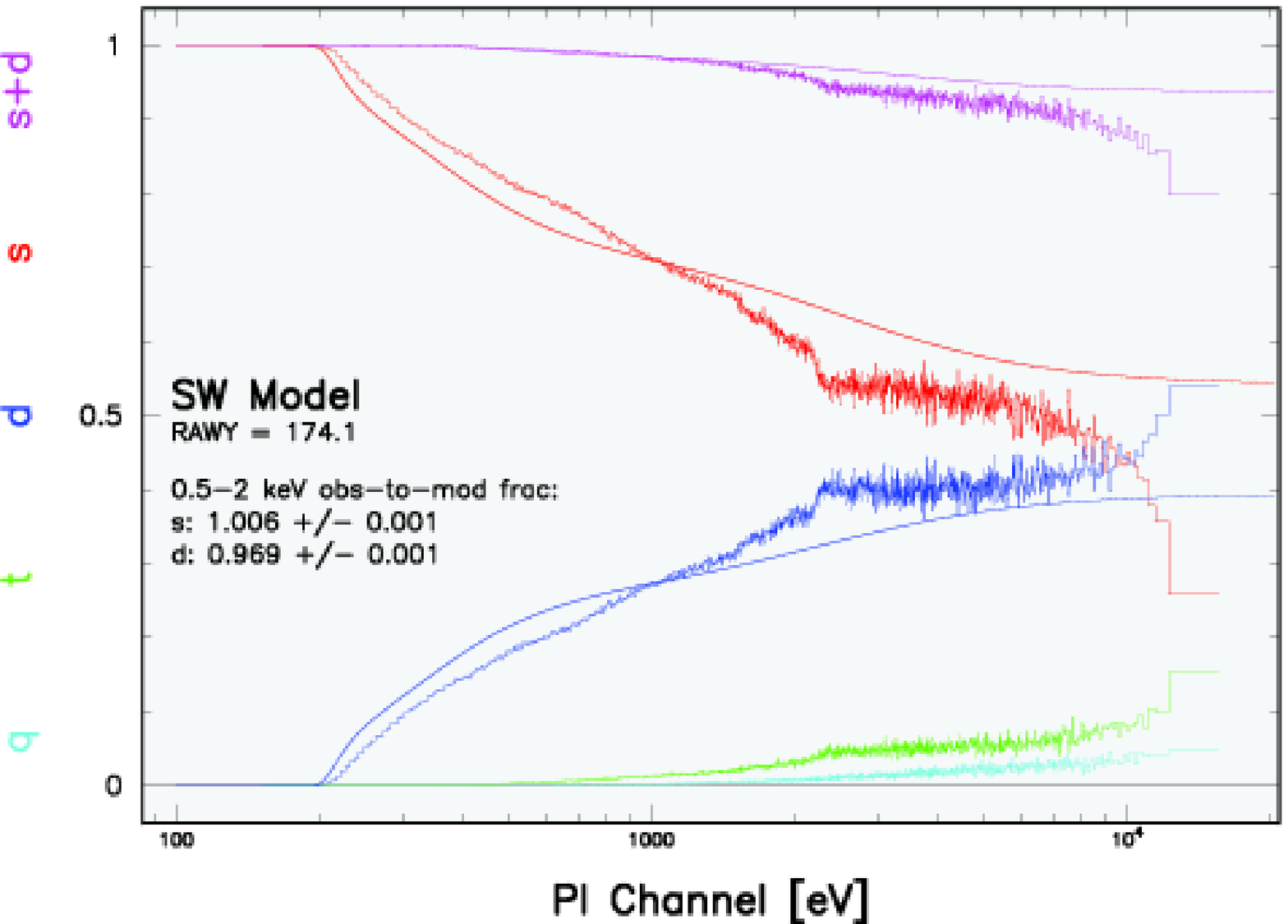}
\vskip 0.5in
\caption{Pile-up measurement from the SAS task EPATPLOT for MKN 421 Obs. ID: 0099280101.  The expected fractions of pixel events are the solid lines and the measured pixel events are the histograms (s,d,t,q stand for single, double, triple, and quadruple events respectively). In the range 0.75 to 10 keV we see that a larger than expected fraction of double events (third histogram from top, dark blue in electronic edition) is measured, indicating that pile-up is present in this source.  [See the electronic edition of the Journal for a color version of this figure.]
\label{epatplot}}
\end{figure}

Some of the observations in our sample are affected by pile-up.  Pile-up occurs when more than one X-ray photon arrives in a pixel before the pixel is read out by the CCD, making it difficult to distinguish one high energy photon from two lower energy photons.  Pile-up can also occur when photons striking adjacent pixels are confused with a single photon that deposits charge in more than one pixel.  Depending on how many pixels are involved, this is called a single-, double-, triple-, or quadruple- pixel event.  
The SAS task EPATPLOT measures the pile-up in an observation and the results for our target with the highest count rate, MKN 421 (Table 1), are shown in Figure~\ref{epatplot}.  
When we compare the expected fractions of pixel events (solid lines) with those actually measured in the data (histograms) for the range 0.75 to 10 keV we see that a larger than expected fraction of double events (third histogram from top, dark blue in electronic edition) is measured as well as a larger fraction of triple and quadruple events (bottom two histograms), although to a lesser degree, while single events (second histogram from top) are lower than expected, indicating the presence of pile-up.  Pile-up leads to a general reduction in the mean count rate as well as a reduction in the magnitude of variations.  We will explore the influence of pile-up on our data in more detail when we discuss structure functions in Section~\ref{sfpileupsection}.

\section{Data Analysis and Results}

\subsection{Wavelet Analysis} \label{waveletanalysis}

\subsubsection{The Continuous Wavelet Transform}
The continuous wavelet transform (CWT) is the inner product of a dilated and translated mother wavelet and a time series $f(t)$, the idea being that the wavelet is applied as a band-pass filter to the time-series.  The continuous wavelet transform maps the power of a particular frequency (i.e. dilation) at different times in translation-dilation space, giving an expansion of the signal in both time and frequency.  Hence, the continuous wavelet transform not only tells us which frequencies exist in the signal, but also when they exist, allowing us to see whether a timescale varies in time.  This is the wavelet technique's advantage over Fourier transforms in detecting quasi-periods.  In addition, the Fourier transform is not suited for detecting quasi-periods since non-periodic outbursts will spread power across the spectrum and windowing will cause power to appear at low frequencies, potentially obscuring quasi-periodic signals.

Throughout this paper, we follow \citet{hug98} and \citet{kel03} and references within.  
In previous studies \citep{hug98,kel03,liu05,kadler06} we have found the
Morlet wavelet
\begin{equation}
     \psi_{Morlet} = \pi^{-1/4} e^{ik_{\psi}t} e^{-|t^{2}|/2},
\end{equation}
with $k_{\psi}=6$ to be an excellent choice. The value of $k_{\psi}$
is a satisfactory compromise between a value small-enough that we
have good resolution of temporal structures, and large-enough that
the admissibility condition is satisfied, at least to machine accuracy
\citep{far92}. The wavelet, being continuous and complex,
permits a rendering in transform space that highlights temporally
localized, periodic activity -- oscillatory behavior in the real part and
a smooth distribution of power in the modulus -- and being progressive
(zero power at negative frequency), is optimal for the study of causal
signals.  We have deliberately avoided any form of weighting, such as that
introduced by \citet{foster} to allow for uneven sampling,
or \citet{johnson} to rescale within the cone of influence,
in order to facilitate our interpretation of the cross wavelet, and to
allow the use of existing methods of significance analysis.  

From this mother wavelet, we generate a set of translated ($t'$) and dilated ($l$) wavelets
\begin{equation}
\psi_{lt^{'}}(t)=\frac{1}{\sqrt{l}} \psi (\frac{t-t^{'}}{l}),  l \in \Re^{+},  t\in \Re
\end{equation}
and we then take the inner product with the signal $F(t)$ to obtain the wavelet coefficients  
\begin{equation}
\label{coefficients}
\widetilde f (l,t^{'})= \int_{\Re}f(t)\psi^{*}_{lt'}(t) dt .
\end{equation}
The wavelet coefficients are later mapped in wavelet space which has as coordinates translation and dilation, and so periodic behavior shows up as a pattern over all translations at a specific dilation.  

\begin{figure}
\figurenum{2}
%\plotone{full_sine_b_norm.ps}
\plotone{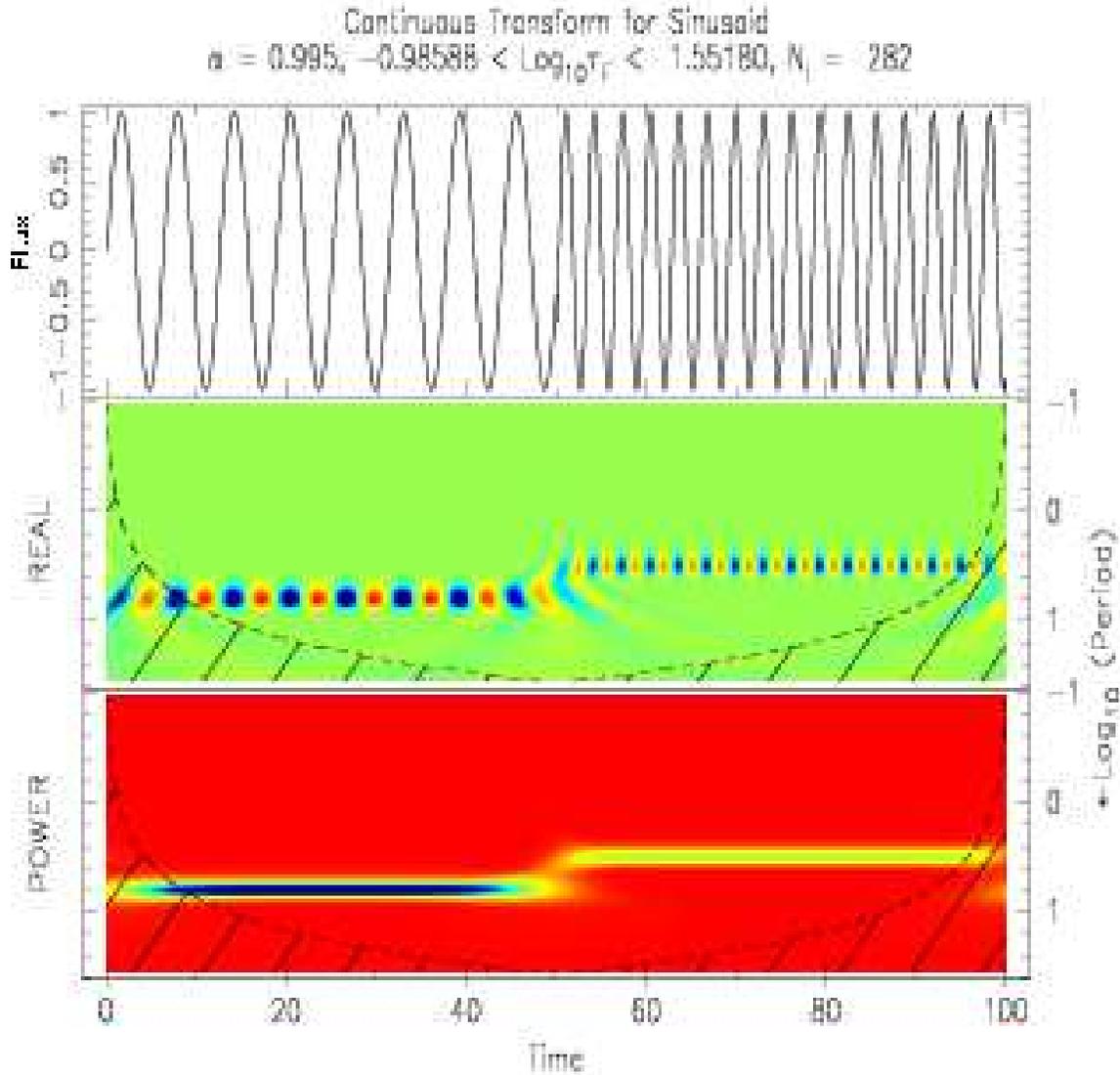}
\vskip 0.5in
\caption{Continuous wavelet transform for a sinusoid of varying frequency (top panel). Periodic behavior shows up as a pattern over all translations (y-axis) at a specific dilation (x-axis) in both the real part of the transform (middle panel) and the power (bottom panel).  Here there are signals at 3s and 6s.  Also given in the figure header are $\alpha$ (the autocorrelation of the signal), the range of dilations explored, and $N_{l}$ (the number of dilations used).  The hatched area represents the cone of influence. [See the electronic edition of the Journal for a color version of this figure.]
\label{sinecwt}}
\end{figure}

By way of example, Figure~\ref{sinecwt} shows the real part and the power of the continuous wavelet transform (second and bottom panel, respectively) for a sinusoidal signal of varying frequency (top panel).  Here, the real part of the transform shows oscillatory behavior corresponding to the two periodicities of the sinusoidal signal at dilations of 3s and 6s with a break in translation at 50s corresponding to the time where the change in frequency occurs.  The bottom panel in Figure~\ref{sinecwt} shows that the power of the continuous wavelet transform is concentrated at these two frequencies as well.    

The hatched area in both panels of Figure~\ref{sinecwt} represents the cone of influence:  the region where edge effects become important. It arises because discontinuities at the beginning and end of a finite time series result in a decrease in the wavelet coefficient power.  Also shown in the header of Figure~\ref{sinecwt} are the number of dilations used ($N_l$) and the ranges of dilations explored.  We discuss $\alpha$ and the normalization of Figure~\ref{sinecwt} in the following section.

%A more detailed description of the continuous wavelet transform can be found in \citet{far92}.  For other studies using the continuous wavelet transform in this context refer to \citet{hug98} and \citet{kel03}.

\subsubsection{Significance Tests} \label{sigtestsection}

Significance tests can be created for the continuous wavelet transform and here we follow \citet{tor98}.  First, one compares the wavelet power with that of an appropriate background spectrum.

We use the univariate lag-1 auto-regressive [AR(1)] process given by
\begin{equation}
\label{noiseeqn}
x_{n}=\alpha x_{n-1} + z_{n}
\end{equation}
where $\alpha$ is the assumed lag-1 autocorrelation and $z_{n}$ is a random deviate taken from white noise.  Note that $\alpha = 0$ gives a white noise process.  Throughout this paper, we will use ``white noise'' to refer to an AR(1) process with $\alpha = 0$.  Red noise is sometimes used to refer to noise with $\alpha = 1$, however, throughout this paper we apply the term to any non-zero $\alpha$.  

The normalized discrete Fourier power spectrum of this process is
\begin{equation}
\label{fouriereqn}
P_{j}=\frac{1-\alpha^{2}}{1 + \alpha^{2} - 2 \alpha \cos(2 \pi \delta t /\tau_{j})}
\end{equation}
where $\tau_{j}$ is the associated Fourier period for a scale $l_{j}$.  %A white noise process gives $P_{j} = 1$ and a red noise process with $\alpha = 1$ gives $P_{j} = 0$. 
We use the above two equations to model a white noise or red noise spectrum.

\begin{figure}
\figurenum{3}
\plotone{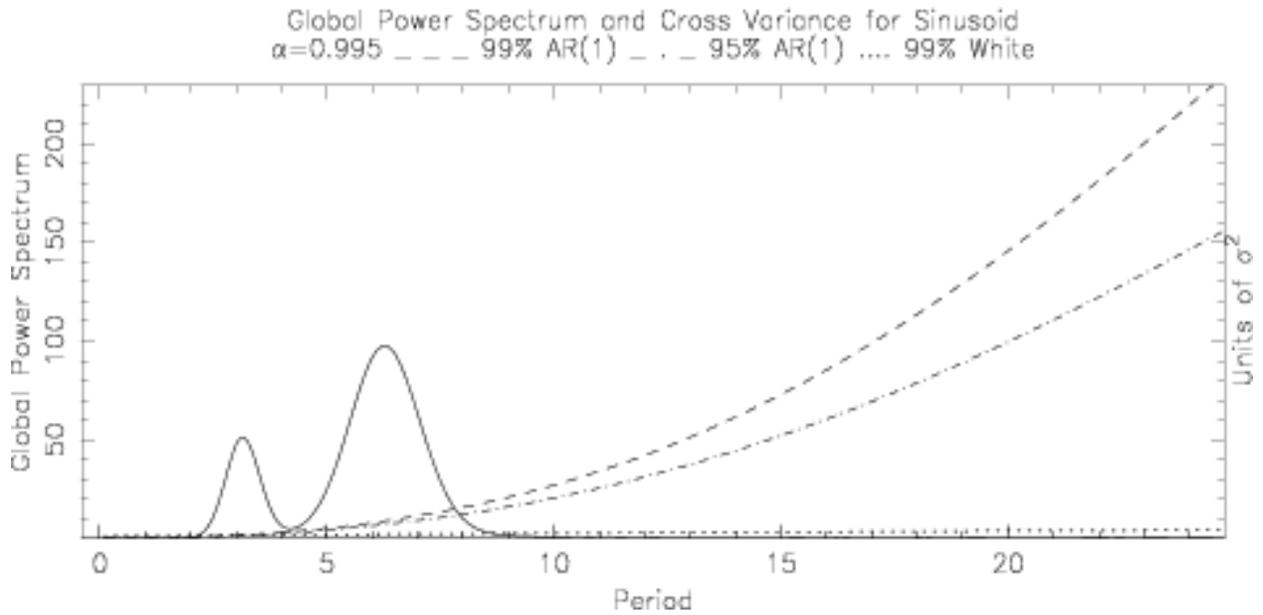}
%\plotone{full_sinegwps.ps}
\caption{Global wavelet power spectrum for a sinusoid of varying frequency. The solid line is the power spectrum of the signal, which is compared to the power spectra of white and red noise random processes (broken lines).  The 3 s and 6 s detections exceed the expected levels of white and red noise at the 99$\%$ significance level.\label{sinegwps}}
\end{figure}

The global wavelet power spectrum (GWPS) is obtained by averaging in time 
\begin{equation}
\widetilde f_{G}^{2}(l_{j})=\frac{1}{N_{j}} \Sigma^{i'_{j}}_{i=i_{j}}|\widetilde f(l_{j},t'_{j})|^{2}.
\end{equation}
Here, $i_{j}$ and $i'_{j}$ are the indices of the initial and final translations $t'_{i}$ outside of the cone of influence at a given scale $l_{j}$.  $N_{j}$ is the number of translations $t'_{i}$ outside the cone of influence at that scale.  
Assuming a background spectrum given by Eqn.~\ref{fouriereqn} we estimate the autocorrelation coefficient ($\alpha$) by calculating the lag-1 and lag-2 autocorrelations, $\alpha_{1}$ and $\alpha_{2}$.  The autocorrelation coefficient is then estimated as $\alpha = (\alpha_{1} + \sqrt{\alpha_{2}})/2$.  The background spectrum $P_{j}$ then allows us to compute the confidence levels.  It is assumed that the time series has a mean power spectrum given by Eqn.~\ref{fouriereqn} and so if a peak in the wavelet power spectrum is significantly above this background spectrum, then the peak can be assumed to be a true feature.  If the values in the time series $f(t)$ are normally distributed, we expect the wavelet power $|\widetilde f|^{2}$ to be $\chi^{2}$ distributed with two degrees of freedom ($\chi^{2}_{2}$).  The square of a normally distributed variable is $\chi^{2}$ distributed with one degree of freedom and the second degree of freedom comes from the fact that both the real and imaginary parts of the complex $\widetilde f$ are normally distributed.  For example, to determine the 95$\%$ confidence level, one multiplies the background spectrum (Eqn.~\ref{fouriereqn}) by the 95th percentile value for $\chi_{2}^{2}$.  In Figure~\ref{sinegwps} we show the GWPS of a time series along with 99$\%$ and 95$\%$ confidence level for a red noise process and 99$\%$ confidence level for a white noise process.

The distribution for the local wavelet power spectrum is
\begin{equation}
\frac{|\widetilde f(l_{j},t'_{i})|^2}{\sigma^{2}} \Rightarrow P_{j}\frac{\chi^{2}_{\nu}}{\nu}
\end{equation}
where the arrow means ``distributed as," $\sigma^{2}$ is the variance, and $\nu$ is the number of degrees of freedom, which is two here.  The indices on the scale $l$ are $j$=1,2,...,$J$ where $J$ is the number of scales, and the indices on the translation $t'$ are $i$=1,2,...,$N_{data}$. We evaluate this equation at each scale to get 95$\%$ confidence contour lines and in this paper our continuous transforms are normalized to the 95$\%$ confidence level for the corresponding red noise process.  Doing this allows one to see the strength of the wavelet coefficients relative to the 95$\%$ confidence level of a red noise process.

\subsubsection{The Cross-Wavelet Transform} \label{xwtsection}

Although the continuous wavelet transform is useful in examining how a time series varies in time and scale, it does not tell us how the time series varies in dilation over a range of scales when assigning a characteristic timescale.  Since a quasi-periodic signal has no unique dilation we use the cross-wavelet transform (XWT) which filters out noise and reveals the QPO more clearly.  Here we use the XWT introduced by \citet{kel03}. 

After the continuous transform identifies that a periodic pattern exists in the data, the dilation that characterizes this period is obtained from the global wavelet power spectrum and is used to create a sinusoidal mock signal.  The continuous wavelet transform coefficients of the data signal $f_{a}(t)$ are then multiplied by the complex conjugate of the continuous transform coefficients of a mock signal $f_{m}(t)$.  The results are  mapped out in wavelet space and analyzed for a correlation.  

The cross-wavelet transform takes the form
\begin{equation}
\widetilde f_{c} (l,t')= \widetilde f_a(l,t')\widetilde f^{*}_m(l,t')
\end{equation}
where the continuous wavelet coefficients $\widetilde f_a$ and $\widetilde f_m$ are given by Equation~\ref{coefficients}.  

\begin{figure}
\figurenum{4}
%\plotone{full_sinecross.ps}
\plotone{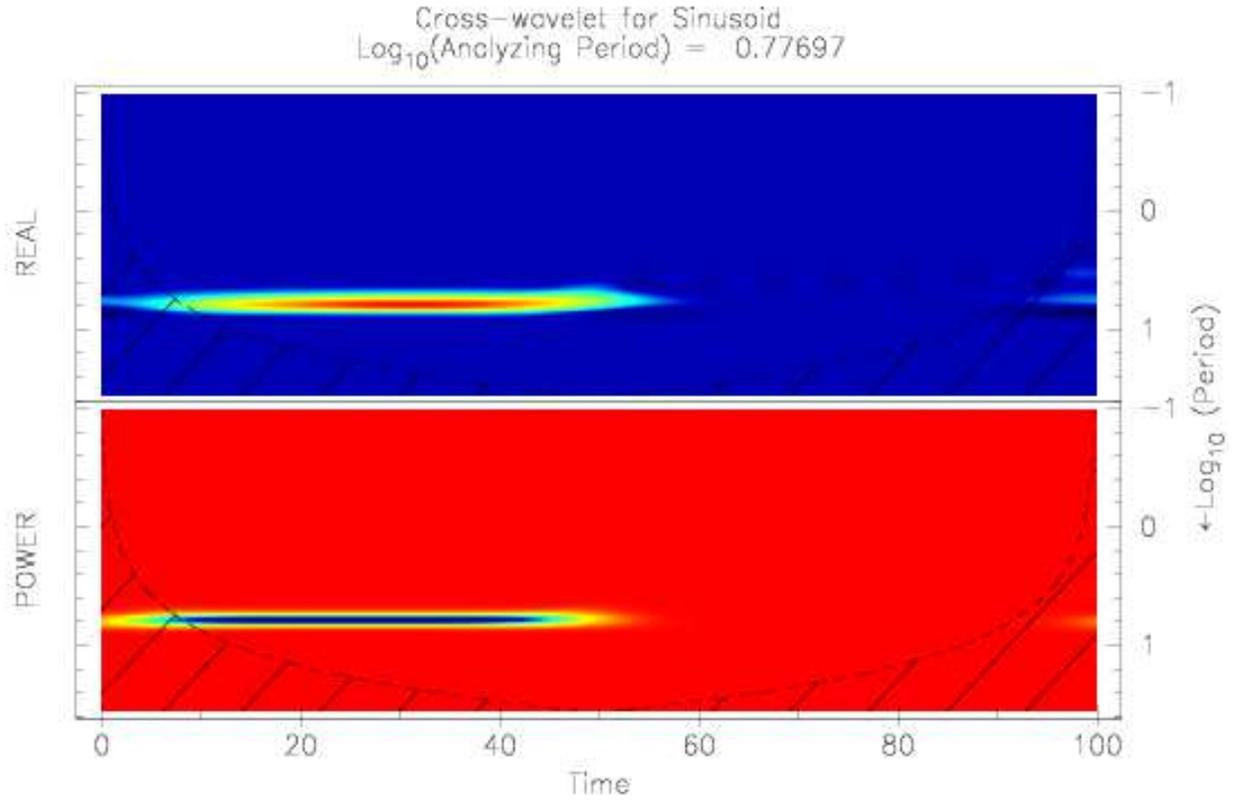}
\vskip 0.5in
\caption{Cross-wavelet transform for a sinusoid.  We use the cross-wavelet to compare a mock sinusoidal signal with a period of 6 s against the sinusoid with varying frequency from Fig.~\ref{sinecwt}.  The concentration in the real and power plots show that a 6s period exists in the first half of the data.  [See the electronic edition of the Journal for a color version of this figure.]\label{sinecross}}
\end{figure}

Figure~\ref{sinecross} shows the cross-wavelet for the same sinusoidal signal of varying frequency used in Figure~\ref{sinecwt}.  The mock signal was calculated using the 6s period found in the wavelet power spectrum (see Fig.~\ref{sinegwps}) and as the concentrations in the real and power panels of Figure~\ref{sinecross} show, the cross-wavelet finds that this 6s period exists in the first half of the time series, illustrating the cross-wavelet's ability to highlight a QPO.

The reader may refer to \citet{kel03} for a full review of the cross-wavelet technique used here.

\subsection{Structure Function Analysis} \label{sfanalysissection}

Since the global wavelet power spectrum compares the observed signal to expected levels of red noise and white noise, we created structure functions (SFs) for each of our observations to see which noise process dominates the signal at different times.  

A structure function calculates the mean deviation of data points, providing an alternate method of quantifying time variations.  Here we use a structure function of the first-order \citep{sim85}:
\begin{equation}
SF(\delta t) = <[F(t) - F(t + \delta t)]^{2}>
\end{equation}
where $F(t)$ is the flux at time $t$ and $\delta t$ is a time lag.  The slope $\alpha$ of the SF curve in $log(SF)-log(\delta t)$ space depends on the noise processes underlying the signal, giving us an indication of the nature of the process of variation.  If $\alpha = 1$ red noise dominates, and for flatter slopes of $\alpha = 0$ Poisson photon noise is significant.  A plateau at short time lag is due to measurement noise.  The transition from plateau to power-law in the structure function curve determines where the dominant underlying noise process changes in the object.  The point of turnover from power-law to plateau at longer time lags corresponds to a maximum characteristic timescale.

\subsubsection{Effects of Pileup} \label{sfpileupsection}

\begin{figure}
\figurenum{5}
\includegraphics[scale=0.45]{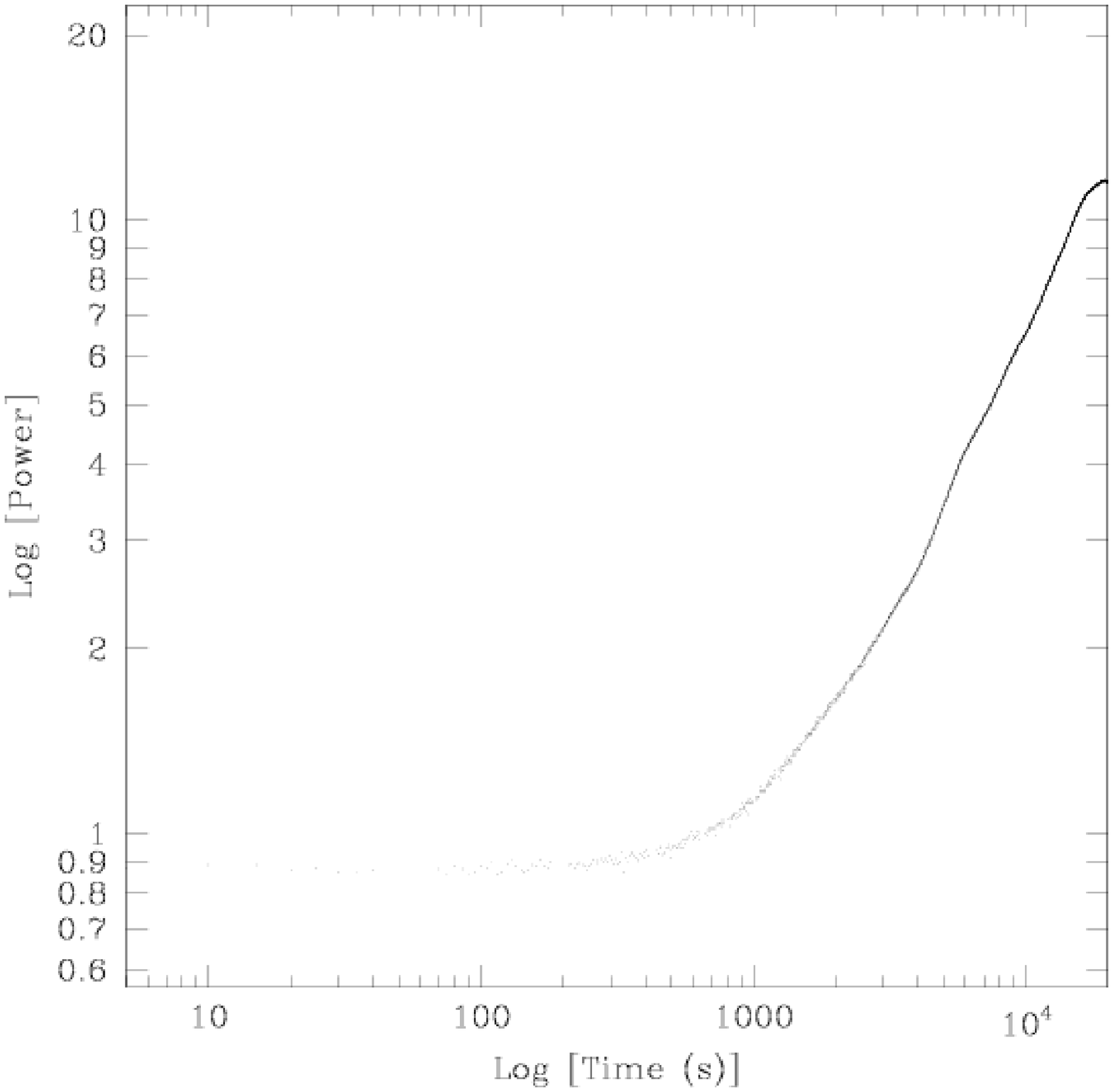}
\includegraphics[scale=0.45]{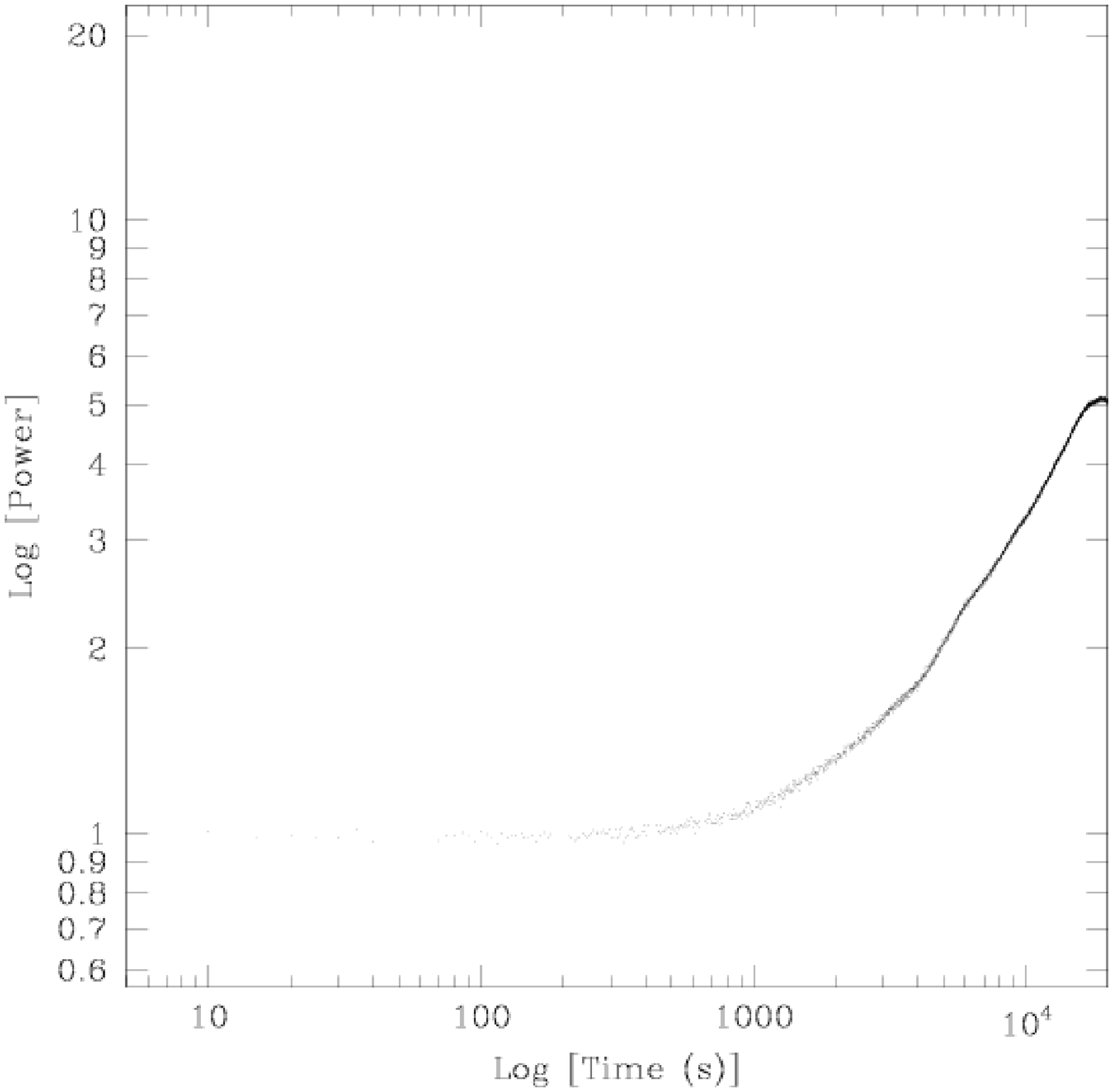}
\caption{Effect of pile-up on structure functions of MKN 421 Obs. ID: 0099280101.  The flat portion of the structure function curve affected by pile-up (left panel) falls below $log(SF)=1$, which corresponds to Poisson photon noise.  In the right panel the central core of the source has been removed (i.e. pile-up is  greatly reduced) and the SF curve is then at the expected value for Poisson photon noise (right panel).\label{structpileup}}
\end{figure}

We measure the presence of pile-up in our observations by using the SAS task EPATPLOT and find that the majority of our sources show varying degrees of pile-up.  For example, as previously shown in Section~\ref{obssection}, MKN 421 Obs. ID: 0099280101 has a modest amount of pile-up (see Figure~\ref{epatplot}).  In the structure function of this observation (left panel, Figure ~\ref{structpileup}), the flat portion of the structure function curve should have a value of $log(SF)=1$, which corresponds to the Poisson photon noise inherent in the photon statistics.  However, here it falls below the Poisson photon noise level.  To remove the pile-up we exclude the central core of the source in the event file since pile-up is more likely to occur here.  For this subtracted data, the EPATPLOT output indicates that there is no pile-up and the SF curve is then at the expected value for Poisson photon noise (right panel, Figure~\ref{structpileup}).  Pile-up affects the SF because it lowers the overall count rate and thereby Poisson photon noise is underreported.  

We correct for pile-up in the rest of our data by adding a fixed value to $log(SF)$, moving the flat part of the structure function curve up to 1.  All of our observations had less than 5$\%$ pile-up except for PKS 2155-304 Obs. ID 124930301 (6.5$\%$) and both observations of MKN 421 ($\sim 10\%$).  Overall, the percentage of pile-up in our sample increases with the number of counts except for NGC 4151 Obs. ID 112830201 which is 5$\%$ piled-up and has an average of only 25 counts.

\subsection{Results}

\subsubsection{Wavelet Analysis Results} \label{wavelet_results}

\begin{figure}
\figurenum{6}
%\plotone{full_3C273_clean_100s_paper_c.ps}
\plotone{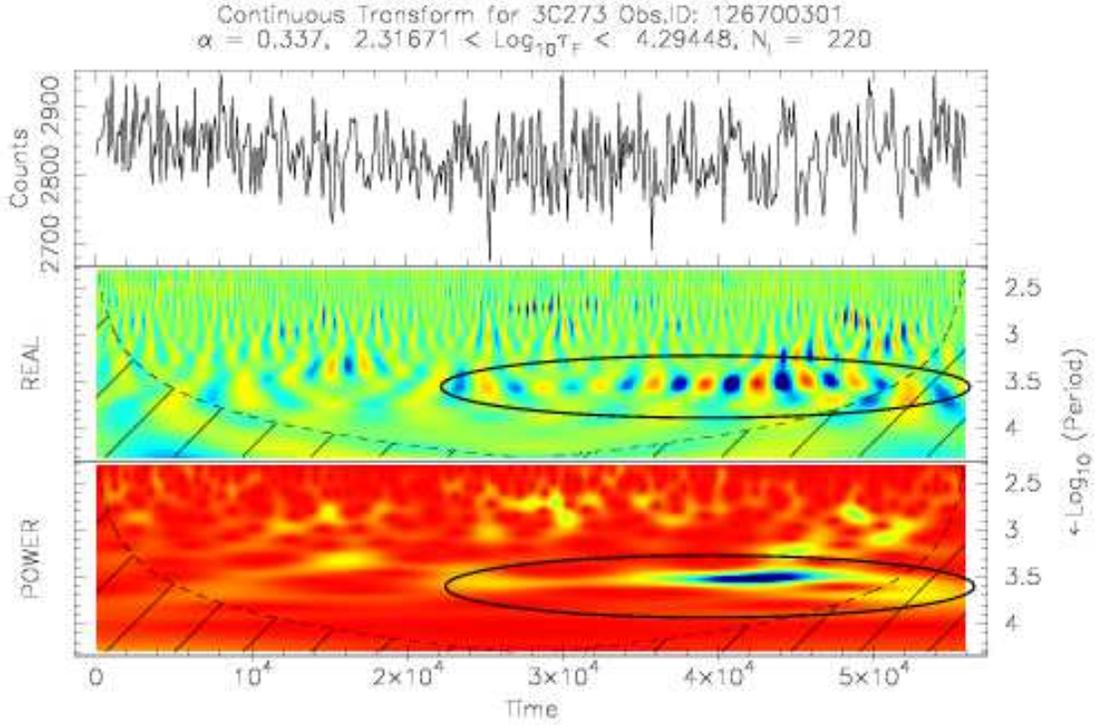}
\vskip 0.5in
\caption{Continuous wavelet transform for 3C 273 Obs. ID: 126700301.  The real part of the transform shows an oscillatory behavior at 3.3 ks.  There is also a concentration in the power plot at 3.3 ks.  The 3.3 ks signal is circled in the both panels.    The light curve (top panel) is binned up from 5s to 100s for clarity.  The $\alpha$ for the unbinned 5s data is 0.14.  The hatched area is the cone of influence: the region where edge effects become important  [See the electronic edition of the Journal for a color version of this figure.]\label{3ccwt}}
\end{figure}

Of the observations that we analyzed, only one showed a quasi-period of
interest (at 3.3 ksec), and this occurred in an observation of 3C 273
(ID 126700301).  The continuous wavelet transform result for this observation is shown in Figure~\ref{3ccwt} with the quasi-period circled in the real and power plots (second and third panel, respectively).  One can see that the quasi-period appears in the last two-thirds of the observation.  In the real plot, the concentrations match up with peaks in the light curve, and the power is concentrated at $4.2\times10^{4}$ s.  The wavelet is sampled with 220 dilations ($N_{l}$) ranging between $\sim 207.2$ s and $2.3\times10^{4}$ s.  We note that the data in Figure~\ref{3ccwt} are binned from 5 s to 100 s for clarity and that we only show the first 56 ks due to background flaring at the end of the observation.  We note that including the periods with background flaring does not change our results.  The $\alpha$ found from autocorrelation analysis for the unbinned data is 0.14 and this value is used to reach the conclusions in this paper.

\begin{figure}
\figurenum{7}
%\plotone{full_3C273_clean_paper_b.ps}
\plotone{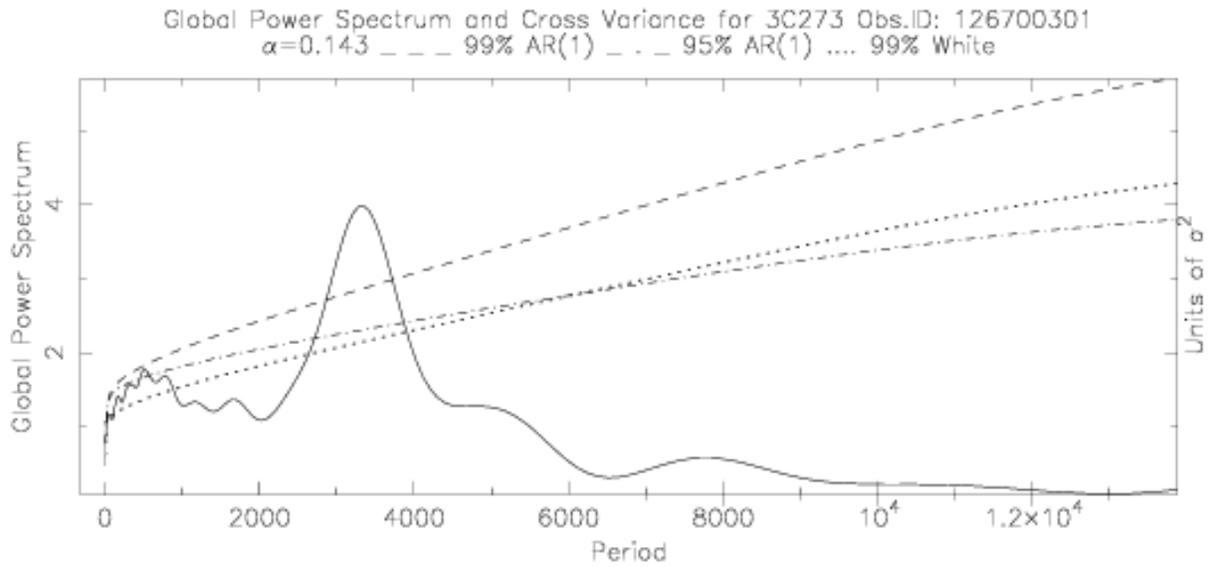}
\caption{Global wavelet power spectrum for 3C 273 Obs. ID: 126700301. The solid line is the power spectrum of the signal, which is compared to the power spectra of white and red noise random processes (broken lines).  The 3.3 ks detection exceeds the expected level of red noise with a 99.979$\%$ probability of detection.  \label{3cgwps}}
\end{figure}

The 3.3 ks quasi-period is also evident in the Global Wavelet Power Spectrum (GWPS, Figure~\ref{3cgwps}), which is calculated by summing up the wavelet power spectra at all times.  In searching for quasi-periodic
behavior we excluded time scales above 25$\%$ of the time series length,
where, using spectral methods, too few periods to provide a convincing
result would be present, and where the cone of influence becomes important
for the wavelet coefficients. On short time scales, experience has shown
that sources often exhibit a broad distribution of power, with local maxima
not well-separated from the mean power level.  We selected a lower bound
for our search, by visual identification of such behavior in the GWPS,
in conjunction with a concomitant change in behavior of the SF.  The solid line in Figure~\ref{3cgwps} is the power spectrum of the signal, which is compared to the power spectrum of white and red noise random processes (broken lines).  One can see that the 3.3 ks detection exceeds the expected levels of white and red noise at the 99$\%$ significance level, i.e. the probability of the detection is higher than 99$\%$ of the noise random processes (the significance of this signal is 99.979$\%$ relative to red noise with $\alpha = 0.14$).  The origins of the white and red noise power spectra were discussed in Section~\ref{sigtestsection}.  

\begin{figure}
\figurenum{8}
%\plotone{full_3C273_clean_crosswav.ps}
\plotone{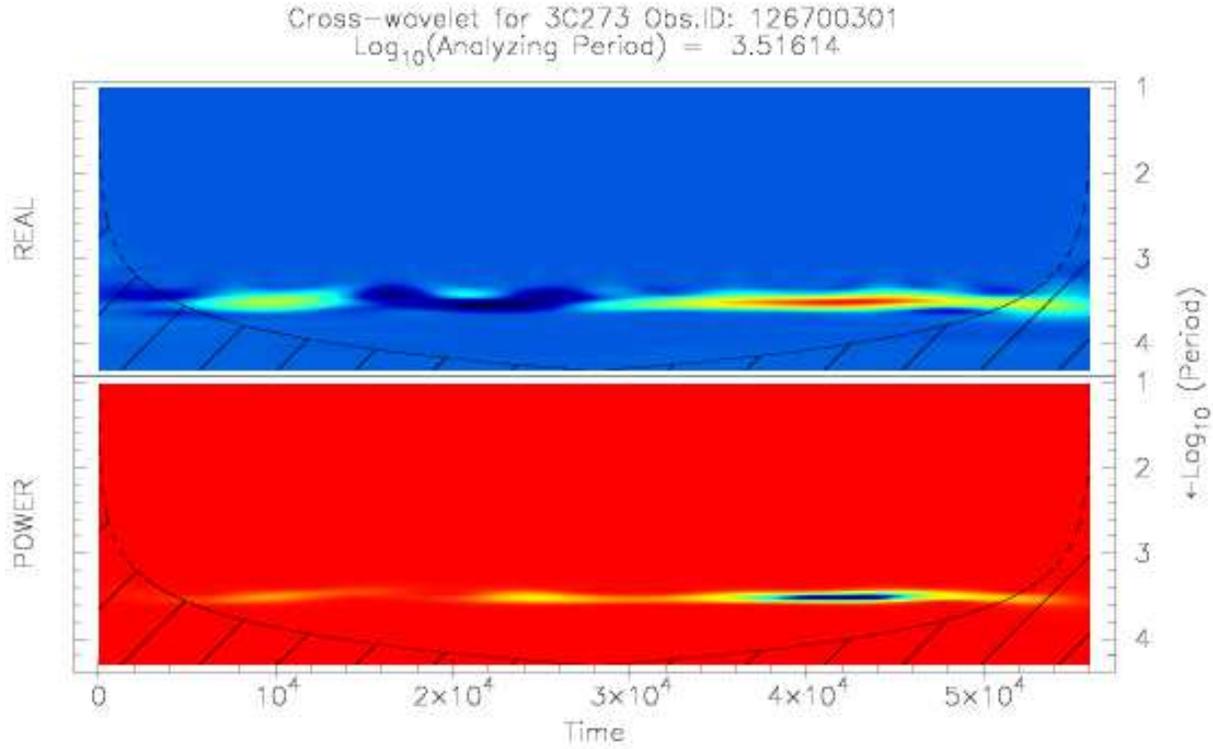}
\vskip 0.5in
\caption{Cross-wavelet transform for 3C 273 Obs. ID: 126700301.  We use the cross-wavelet to compare a mock sinusoidal signal with a period of 3282 s with the observation.  The concentration in the real and power plots show that a $\sim$3300s period exists in the data.  [See the electronic edition of the Journal for a color version of this figure.]\label{3ccross}}
\end{figure}

The cross-wavelet analysis for 3C 273 (Fig.~\ref{3ccross}) supports the conclusion that a period of ~3.3 ks is indeed present.  Here, the XWT (see Section~\ref{xwtsection}) compares a mock sinusoidal signal with a period of 3282 s with the 3C 273 light curve.  The concentration in the cross-wavelet transform shows that the 3.3 ks signal is present throughout the observation.   As one can see, by comparing the
crosswavelet signals in juxtaposed bands, the 3.3 ks periodicity can be traced over the entire interval.  In the CWT
(Figure~\ref{3ccwt}) the 3.3 ks signal is particularly strong at late times,
and so, due to the limited dynamic range of the rendering, is not evident
early in the time interval in that figure.

\begin{figure}
\figurenum{9}
\plotone{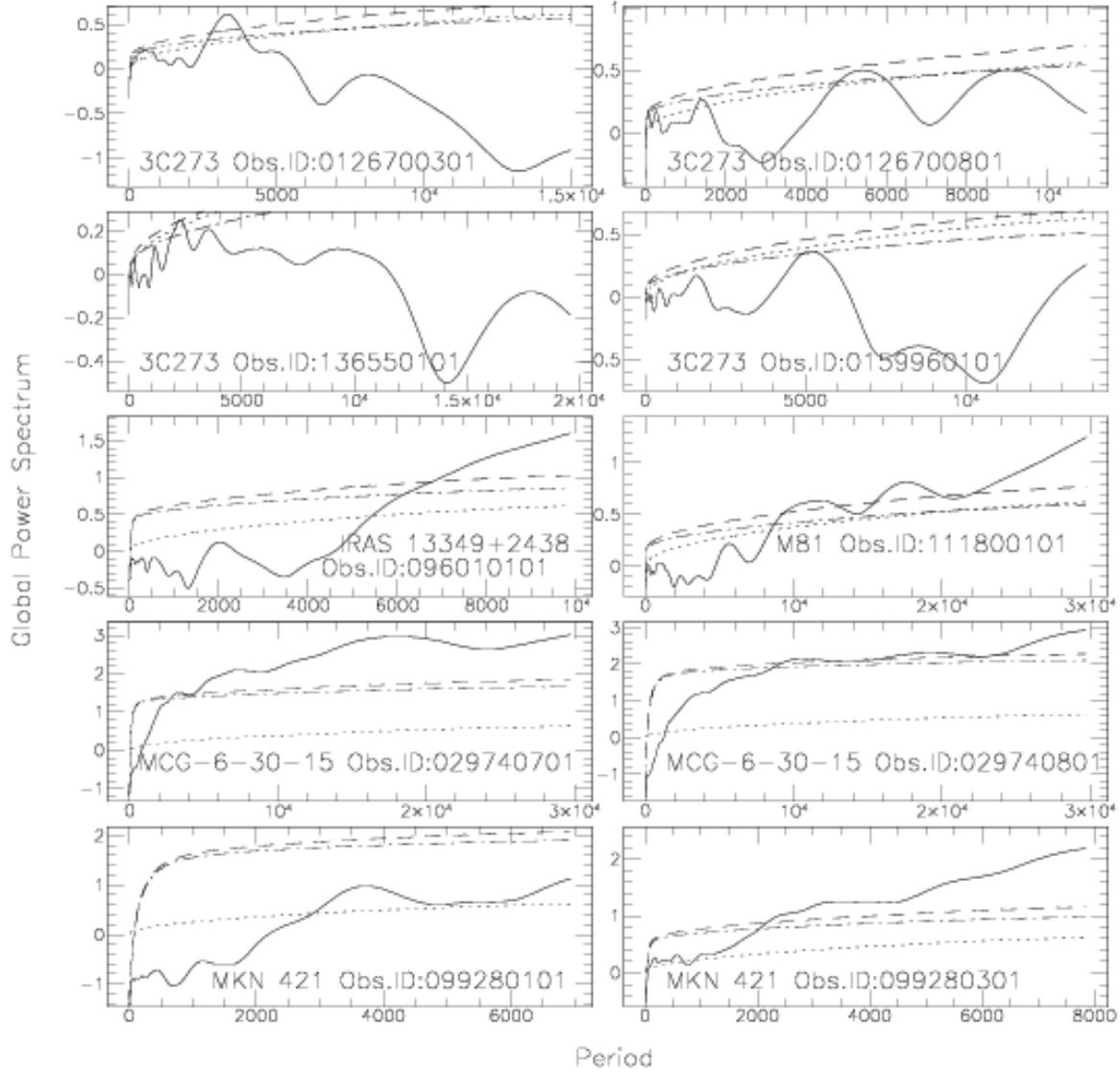}
%\plotone{gwpsall1.ps} 
\caption{Global wavelet power spectra for all observations in our sample with log(Global Power Spectrum).  The period is in seconds. \label{gwpsall1}}
\end{figure}

This periodicity is not detected in the other three observations of this object.  In the 58 ks (Obs. ID 159960101) and 60 ks (Obs. ID 126700801) observations of 3C 273, there is a signal at 5000s, but it does not rise above the 99$\%$ red noise confidence level (Fig.~\ref{gwpsall1}).  We note that a Fourier analysis of 3C 273 yielded a feature at 3.3 ks, but with a lower significance ($<$ 3$\sigma$) than is found with the wavelet technique.  

We performed Monte Carlo simulations in order to estimate the probability that the wavelet technique would claim a spurious detection.  As a baseline, we created one thousand simulated light curves for Poisson photon noise (Fig. 11) to represent random observational errors i.e. photon counting statistics.  The simulated light curves were 56 ks long with 5 s intervals and we multiplied the mean deviate $z_{n}$ by 40 to produce an average spread in the y-axis of 40 counts to resemble the 3C 273 light curve.  Most of the false detections occur at timescales less than 2000s which corresponds to 3.6 $\%$ the length of the observation and supports our earlier point that one can select the lower limit to search for periodicities by visual identification of broad distributions of power on short time scales in the GWPS.  
On average, the wavelet technique claims a detection (at or
above the significance level reported by the wavelet analysis
for 3C 273) 0.4$\%$ of the time (Fig.~\ref{hist}).
%To simulate red noise, we followed the same procedure that we
%used for the white noise process, but with $\alpha=0.14$, taken
%from an autocorrelation analysis of 3C 273.  As the histogram
%in Figure~\ref{hist} (right panel) shows, we find that there
%is a 1$\%$ chance that the wavelet technique will report a
%periodicity at or above the 99.979$\%$ significance reported
%by the wavelet analysis.

The Monte Carlo simulations suggest a significantly higher rate
of false detections than is implied by the statistics based
on the GWPS. However, they are consistent with the latter
estimates {\it within the margin of error}, given that only
1000 realizations of a time series were generated.  Better simulation statistics could be
achieved by increasing the number of time series realizations
by several orders of magnitude, but devoting time and resources
to this is not warranted.  Visual inspection of the simulated
light curves reveals that they differ qualitatively from the
actual time series: a better correspondence can be achieved
with the addition of randomly distributed Gaussian-profile
bursts of fixed, small amplitude. Evidently, the process under
study is not strictly a stationary, first order one, and the
formal statistical measures of significance should be regarded
as only indicative of the high likelihood of a quasi-periodic
phenomenon in this source. A more detailed analysis, allowing
for nonstationary processes, is beyond the scope of this paper.
While we have performed 19 independent experiments and found only 1 detection
we point out that of our 19 data sets
only 7 have average counts (Table 1) equal to or more than the observation
in which we find the QPO. One cannot expect to see with equal likelihood,
a periodicity of equal strength in these weaker AGN.

We note that independently, the XWT finds evidence for
power throughout the observation at 3.3 ks (Fig. 8).
We measured the 3.3 ks signal strength across the time series from the power plot of Figure 8.  The power of the 3.3 ks signal is $\sim$ 4000 times stronger than shorter and longer dilations, illustrating that the 3.3 ks period is well-constrained.  We also ran the XWT on this time series with analyzing signals of 2.3 ks and 4.3 ks.  The average power of these signals is $\sim$ 2 times less than the average power of the 3.3 ks signal.  This demonstrates that the XWT is picking out a well-defined, persistent signal, and will not misleadingly suggest a signal where there is none.

We did not find any significant detections for the other nine AGN in our sample.  No features had significances that exceeded the 99$\%$ confidence levels for both white noise and red noise processes (see Figs.~\ref{gwpsall1},~\ref{gwpsall2}) and appeared at either too short (i.e. at timescales shorter than 3.6$\%$ the length of the observation) or too long (i.e. at timescales greater than half the length of the observation) a timescale.  Some of the AGN in our sample have been studied before and  previous reports of QPOs exist in the literature.  We will discuss those results in more detail in Section~\ref{discussionprevious}.

%As further evidence of the wavelet technique's robustness, we also analyzed X-ray observations of XTE J1550-564 (see Appendix A).  The wavelet technique results are in agreement with previously published QPOs detected using Fourier methods.  

\subsubsection{Structure Function Results}\label{sfanalysis}

After correcting for pile-up, we subtract a constant level corresponding to Poisson photon noise from the structure functions (Figures~\ref{sfminus1},~\ref{sfminus2}).  The slopes are measured by fitting a power-law to the SF curve using the least-squares method in $log(SF)-log(\delta t)$ space.  Slopes are listed in Table~\ref{sftbl} along with the characteristic time-scales of variability, which were measured by identifying the times of turnover from plateau to power-law and vice versa in the SF curve.  All of our structure functions have a flat plateau at short timescales corresponding to Poisson photon noise, most have a power-law portion, and some have a plateau at long timescales.  We include light curves in Figures ~\ref{lightcurves1}, ~\ref{lightcurves2} for comparison with the structure functions.

The structure functions for all four observations of 3C 273
are shown in the first four panels of Figure~\ref{sfminus1}.
The observation with the 3.3 ks quasi-period (upper left,
Figure~\ref{sfminus1}) is dominated by whitish noise around
3000s, as inferred from its flat slope; however, the SF is
unsuited to quantifying the autocorrelation coefficient
precisely.  Recall that the wavelet analysis finds an
autocorrelation coefficient of $\alpha = 0.14$, relatively
small, and consistent with a flattish structure function. We
note that this observation also has the greatest excess of
such noise above the photon noise, compared to the other three
observations, consistent with this being a unique time series
out of all those analyzed.

\section{Discussion} \label{discussion}

\subsection{Mass Estimates of 3C 273}
There are several mass estimates for 3C 273 obtained from different methods.  One method is reverberation mapping whereby one uses the time lag of the emission-line light curve with respect to the continuum light curve to determine the light crossing size of the broad line region (BLR) and then assumes Keplerian conditions in the broad line region gas motion (i.e. $M_{BH}=v^{2}R_{BLR}/G$) \citep{pet00}.  

Reverberation mapping results based on the optical continuum (i.e. Balmer lines) place the mass of the central black hole in 3C 273 at $2.35^{+0.37}_{-0.33}\times 10^8$ M$_\sun$ \citep{kas00}.  In a different study, \citet{pian} use $Hubble$ $Space$ $Telescope$ UV luminosities to find the broad line region size.  To do so, they derive a relationship between $R_{BLR}$ and UV luminosity using the empirical relationship found by \citet{kas00} between $R_{BLR}$ and the optical luminosity.  \citet{pian} obtain a mass of $4.0^{+2}_{-2}\times 10^{8}M_\sun$ for 3C 273, consistent with the \citet{kas00} value within errors.  In another study, \citet{paltani} look at the strongest broad emission UV lines (Ly$\alpha$ and C IV ) in archival $International$ $Ultraviolet$ $Explorer$ observations and obtain a mass of $6.59^{+1.86}_{-0.9}\times10^9M_\sun$ for the central supermassive black hole in 3C 273.

There are also mass estimates for 3C 273 that do not come from reverberation mapping.  \citet{lia03} find a black hole mass of $2\times 10^7$ M$_\sun$ by generalizing the Elliot-Shapiro relation to the Klein-Nishina regime for 3C 273's gamma-ray flux obtained from EGRET.  Another method is to use the \citet{mclure} correlation between host galaxy luminosity and black hole mass which obtains a mass of $1.6 \times 10^9$ M$_\sun$ with an uncertainty of 0.6 dex \citep{wang}.

\subsection{Underlying Physical Process for the QPO in 3C 273}

If the 3.3 ks quasi-period in 3C 273 represents an orbital timescale originating near a last stable orbit, it implies a central black hole mass of $7.3\times 10^6$ M$_\sun$ for a non-rotating black hole or $8.1\times 10^7$ M$_\sun$ for a maximally rotating black hole.  These numbers agree with the \citet{lia03} mass estimate of $2\times 10^7$ M$_\sun$.  However, these masses are substantially lower than those expected for supermassive black holes.  

The \citet{pian} estimate for the mass of the black hole in 3C 273 at $4.0 \times 10^8$ M$_\sun$ points to an orbital period of $\sim 200$ ks for a last stable orbit of 3$R_S$ and a period of $\sim$16 ks for 0.6$R_S$ for a rotating black hole.  \citet{paltani} estimate a mass for 3C 273 of $6.59 \times 10^{9}$ M$_\sun$, which  points to an orbital period of ~3000 ks for a last stable orbit of 3$R_S$ and a period of ~270 ks for 0.6$R_S$.  The 3.3 ks quasi-period we find here is only about 2-20$\%$ of the \citet{pian} orbital timescale and 0.1-1$\%$ of the \citet{paltani} orbital timescale, suggesting that this X-ray quasi-period is not caused by dynamical motion in the inner accretion disk.  Furthermore, the inverse scaling between frequency and black hole mass yields an expected period of $t \sim 300 M_{BH}/10^{6}M_\sun$ based on the representative of HF QPOs in XRBs, GRO J1655-40 \citep{orosz,remi99,abram}.  Using either the \citet{pian} or \citet{paltani} mass estimates yields a period that is one to two orders of magnitude higher than what we observe.

Previous work has suggested that oscillations can occur in the innermost region of relativistic accretion disks due to their instability against axisymmetric radial oscillations, possibly due to a magnetic field \citep{kato}.  This has been proposed for X-ray binary systems, but the physical mechanism responsible for these oscillations can be applied to other accretion disk systems like AGN.  \citet{perez} analyze modes of oscillation in terms of perturbations of the general relativistic equations of motion of perfect fluids within the Kerr metric.  They look at the case of a thin accretion disk around a Kerr black hole in order to determine black hole mass and angular momentum for different trapped modes.
  
We propose that a g-mode oscillation of $m\geq3$ is responsible for the 3.3 ks quasi-period in 3C 273.  A g-mode (inertial) oscillation can be characterized as a restoring force that is dominated by the net gravitational-centrifugal force.  These modes are the most relevant observationally since they appear to occupy the largest area of the disk and hence should be the most observable trapped modes \citep{perez}.

Equation 5.4 of \citet{perez} shows that the frequency of a quasi-period should be observed at
\begin{equation}
f=714(M_\sun/M)F(a) Hz
\end{equation}
where $a$ is the angular momentum parameter and $M=M_{AGN}$.  For $m=0$ and $a=0$,  $F(0)=1$, while $F(a_{max})=3.443$ where $a_{max}=0.998$.  This gives a mass that is too low.  For $m=3$,
$F(a_{max})\sim~59$ (see Figure 5 of \citet{perez}) and this gives a mass for 3C 273 of $1.4 \times 10^8 M_\sun$. \citet{perez} do not look at modes higher than 3.
%However, \citet{perez} state that any mode greater than zero will not be observable if the disk is viewed face-on.  Assuming the accretion disk is perpendicular to the jet, 3C 273 is believed to be a nearly face-on object.

\subsection{Previously Reported QPOs for AGN in Our Sample} \label{discussionprevious}

\begin{figure}
\figurenum{10}
\plotone{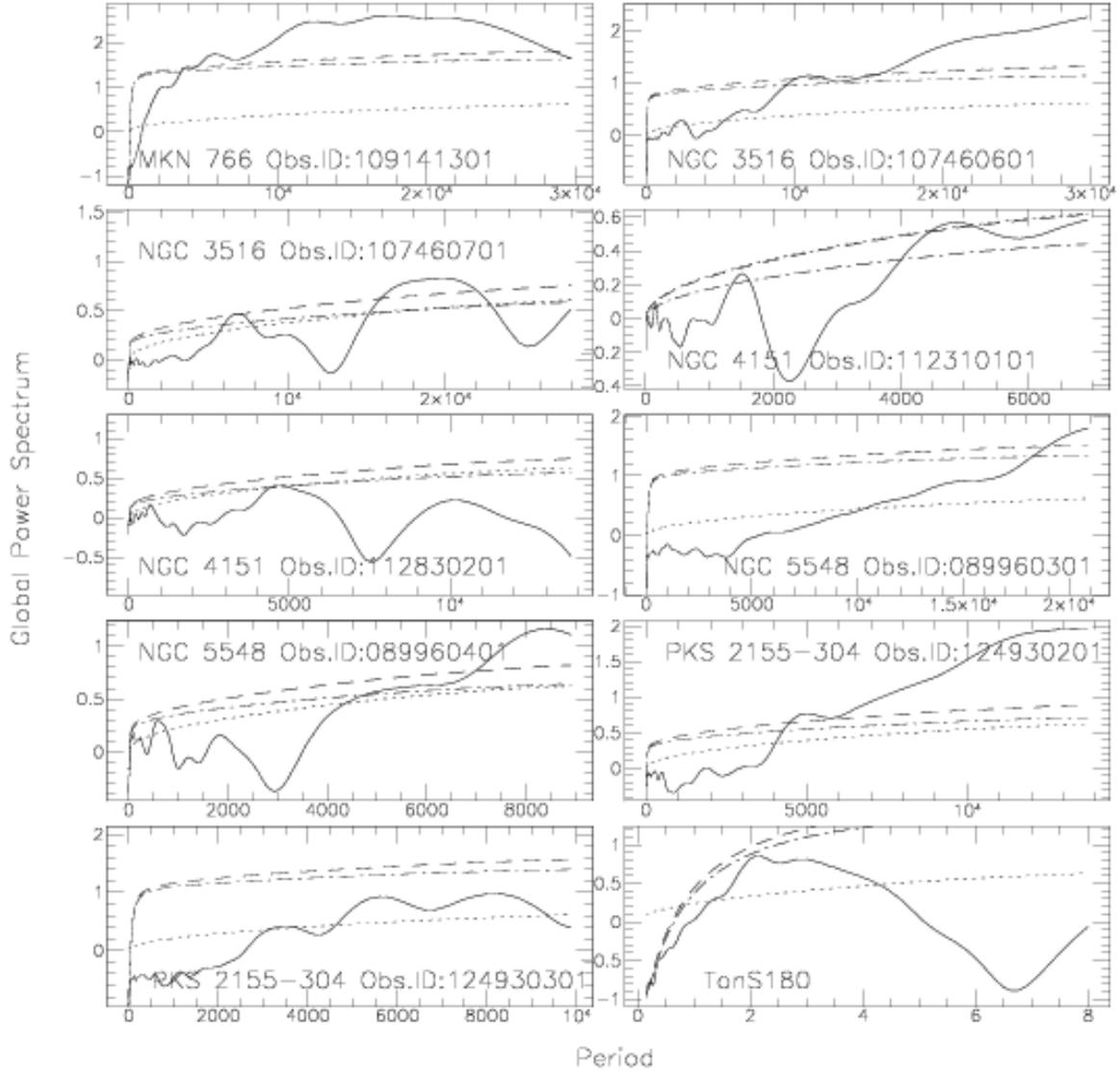}
%\plotone{gwpsall2.ps} 
\caption{Global wavelet power spectra for all observations in our sample with log(Global Power Spectrum).  The period is in seconds except for TonS180 where the period is in days.  TonS180 is not in our sample, but is discussed in Section 4 and so we include it here.\label{gwpsall2}}
\end{figure}

\begin{figure}
\figurenum{11}
\plotone{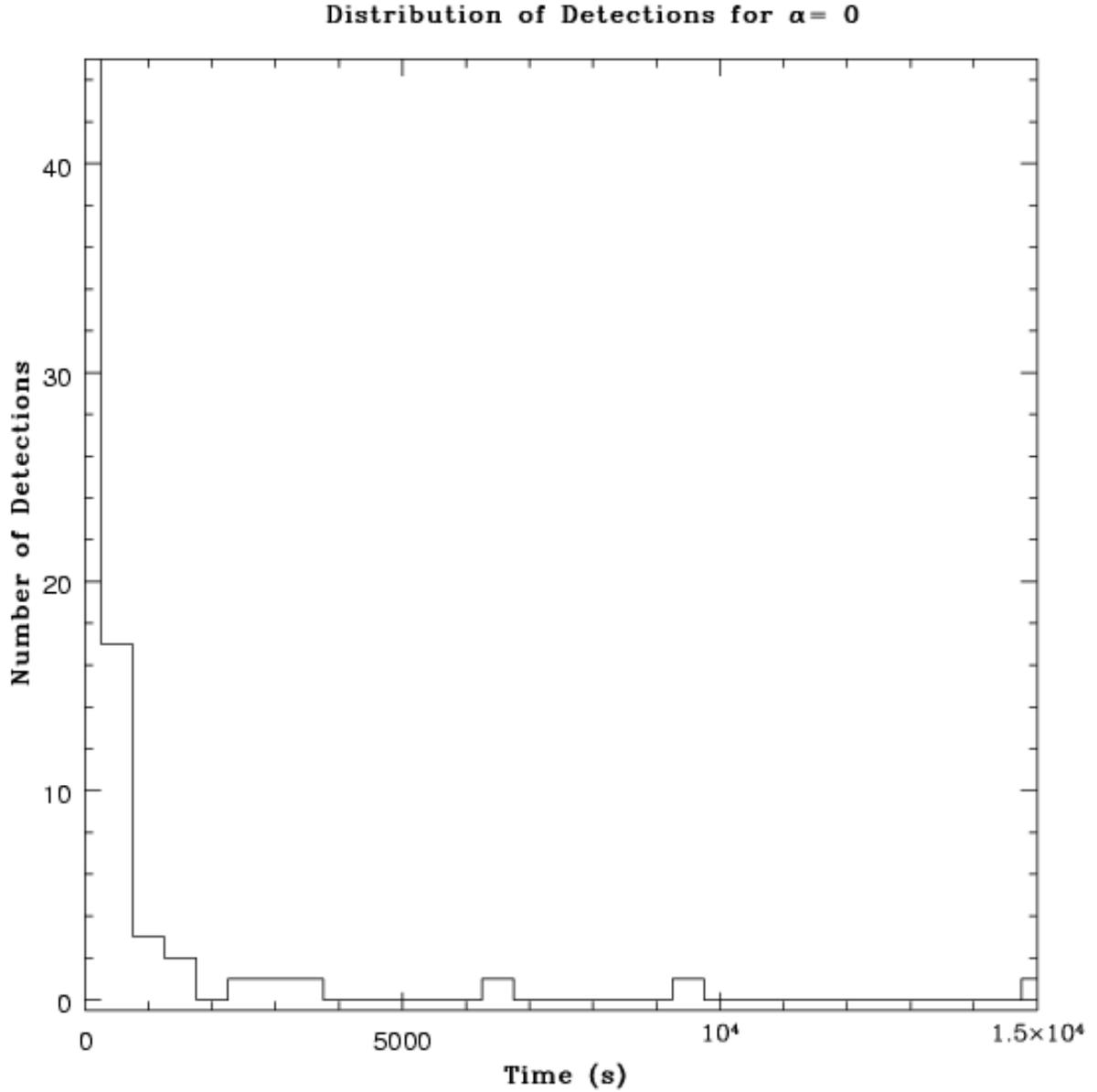}
\caption{Histograms from Monte Carlo significance tests.  
%Histograms are based on one thousand simulated light curves created with parameters set to visually resemble the 3C 273 light curve in which we found the 3.3 ks detection in 3C 273 (see Section ~\ref{wavelet_results} for more details).  
The histogram shows that for simulated light curves
created with Poisson photon noise there
are four false detections (i.e. periodicities at a significance
of $>99.99\%$) at timescales $2000<\tau<1.5\times 10^{4}s$,
translating to a probability of 0.4$\%$.  The y-axis is cutoff
at 45 but rises to 650.  %For $\alpha=0.14$, the histogram
%on the right shows that there are 10 such false detections
%(i.e. periodicities at a significance of $>99.979\%$) and hence
%a 1$\%$ probability of a false detection.  The y-axis is cutoff
%at 45 but rises to 140. The histograms have 500 s bins.
\label{hist}}
\end{figure}

\begin{figure}
\figurenum{12}
\plotone{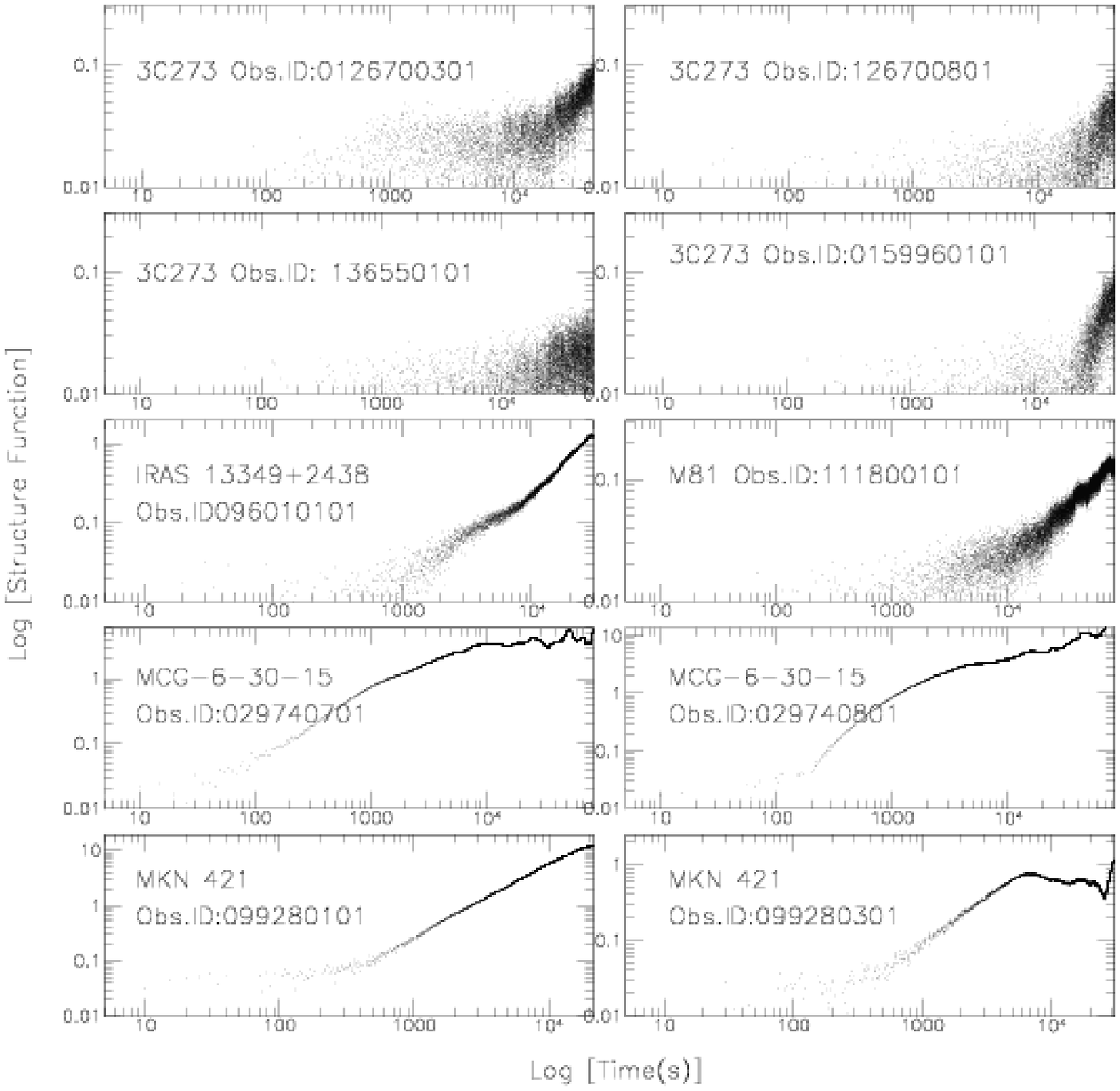}
%\plotone{sfminus_fig_a_paper.ps}
\caption{Structure functions for each object in our sample.  Poisson photon noise has been subtracted.  All of our structure functions have a flat plateau at short timescales corresponding to Poisson photon noise and most have a power-law portion corresponding to red noise. \label{sfminus1}}
\end{figure}

\begin{figure}
\figurenum{13}
\plotone{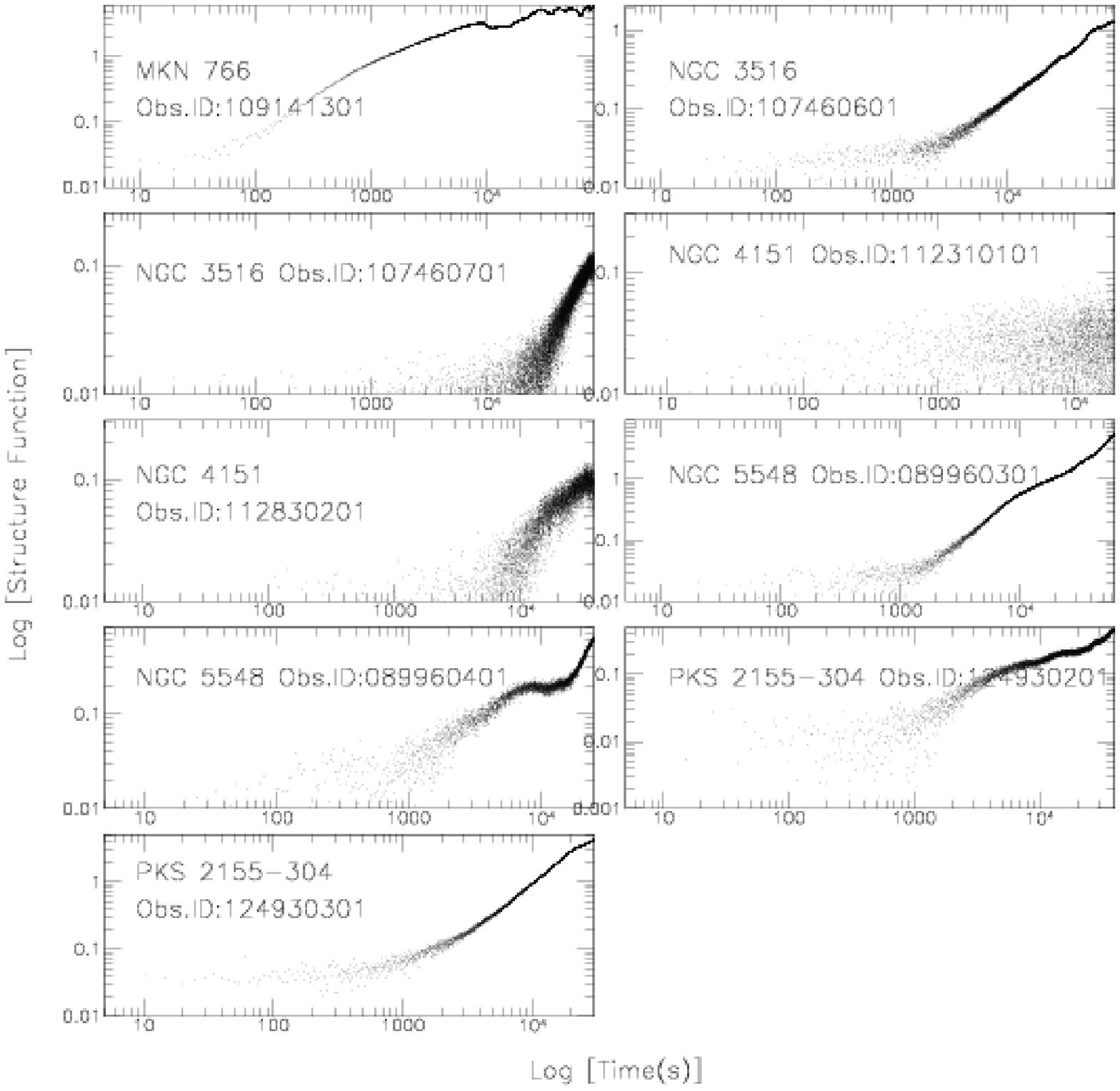}
%\plotone{sfminus_fig_b_paper.ps}
\caption{Structure functions for each object in our sample.  Poisson photon noise has been subtracted. All of our structure functions have a flat plateau at short timescales corresponding to Poisson photon noise and most have a power-law portion corresponding to red noise.\label{sfminus2}}
\end{figure}

\begin{figure}
\figurenum{14}
\includegraphics[scale=0.4]{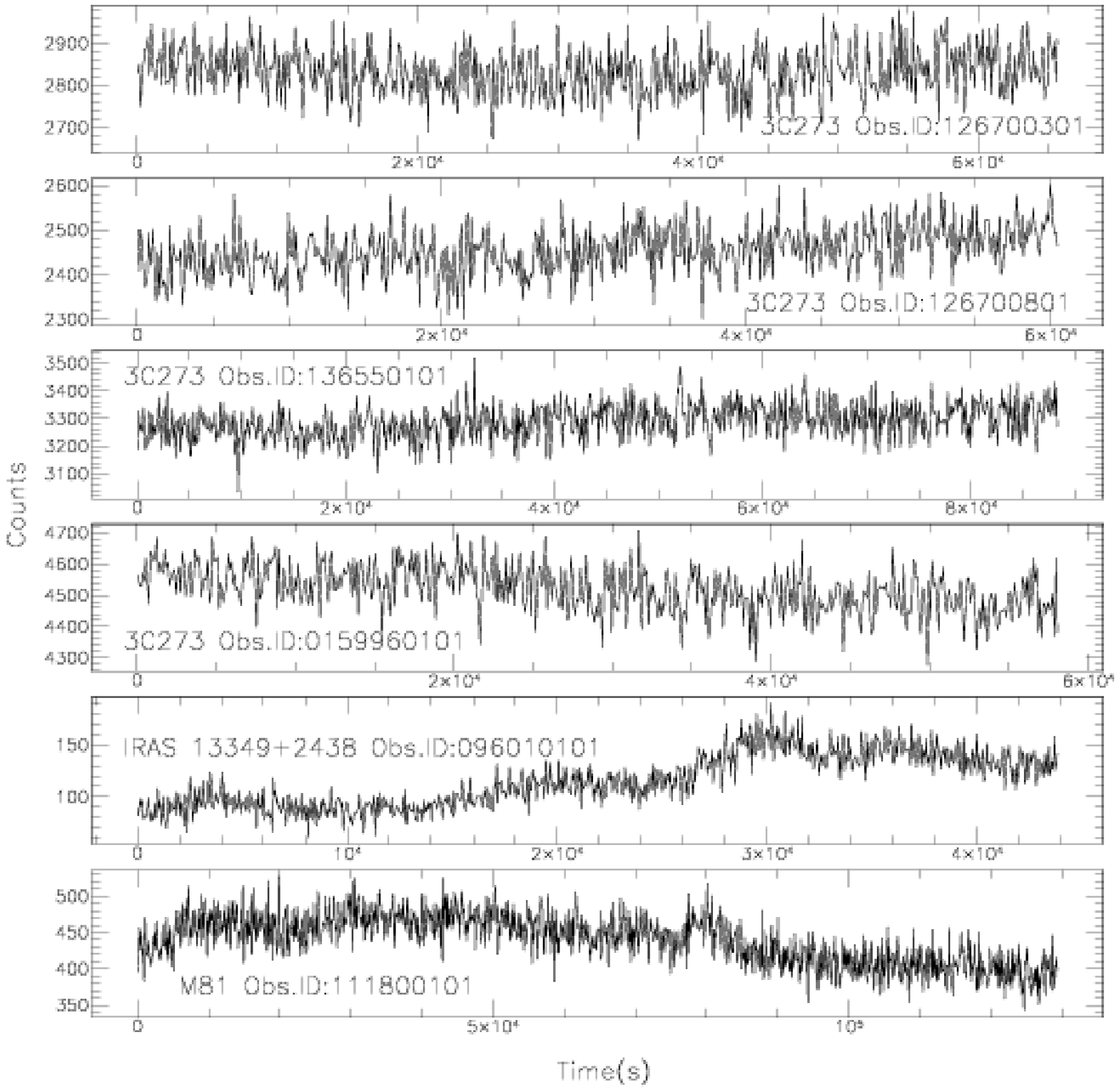}
\includegraphics[scale=0.4]{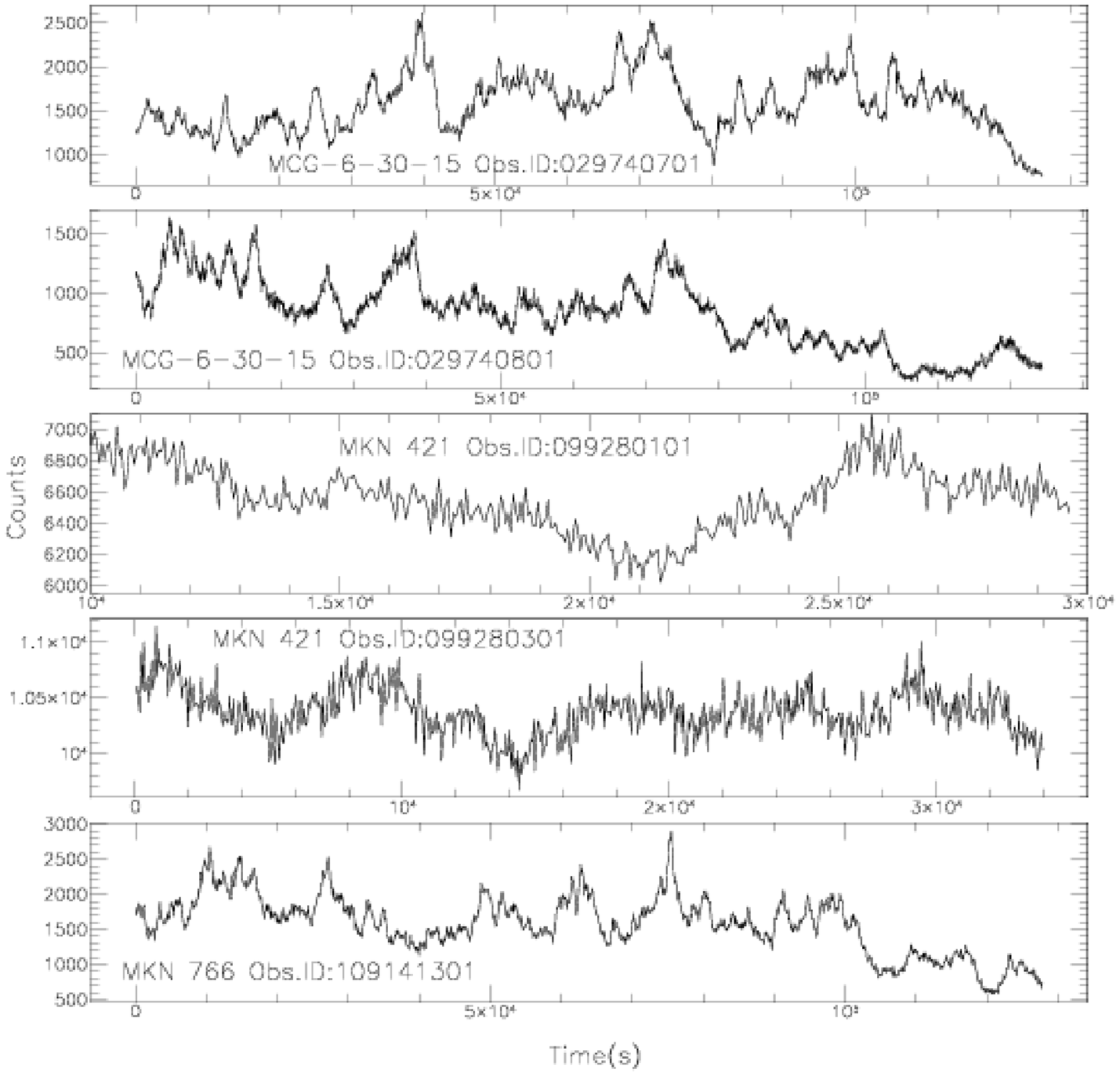}
%\plottwo{figure1.ps} {figure2b.ps}
\caption{Lightcurves for the 10 AGN in our sample.  All observations are in the energy range 0.75 to 10 keV.  Most observations are binned to 100s except for MKN 421, NGC 5548 (Obs ID: 089960401) and IRAS 13349+2438 which are binned to 50s.
\label{lightcurves1}}
\end{figure}

\begin{figure}
\figurenum{15}
\includegraphics[scale=0.4]{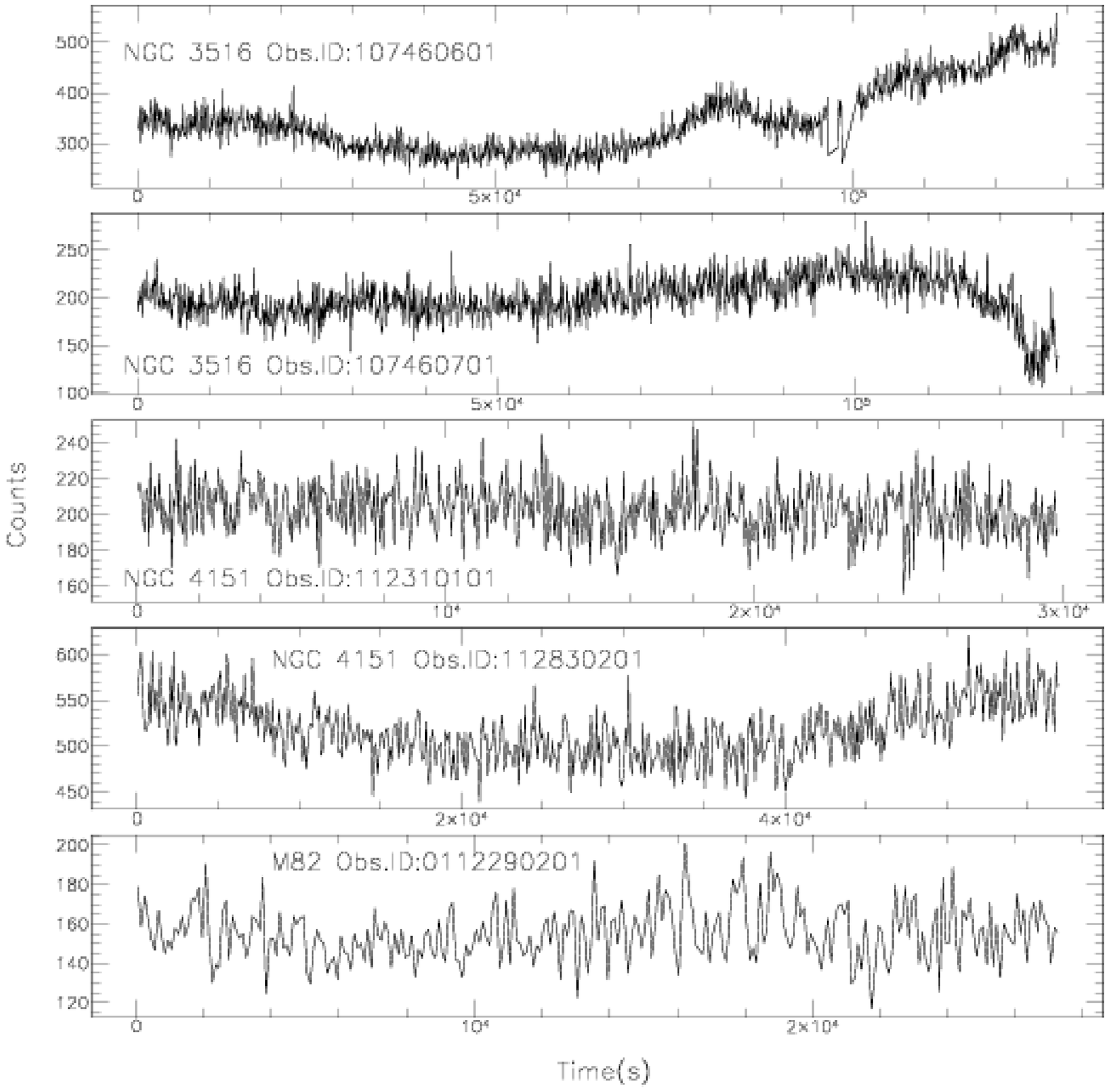}
\includegraphics[scale=0.4]{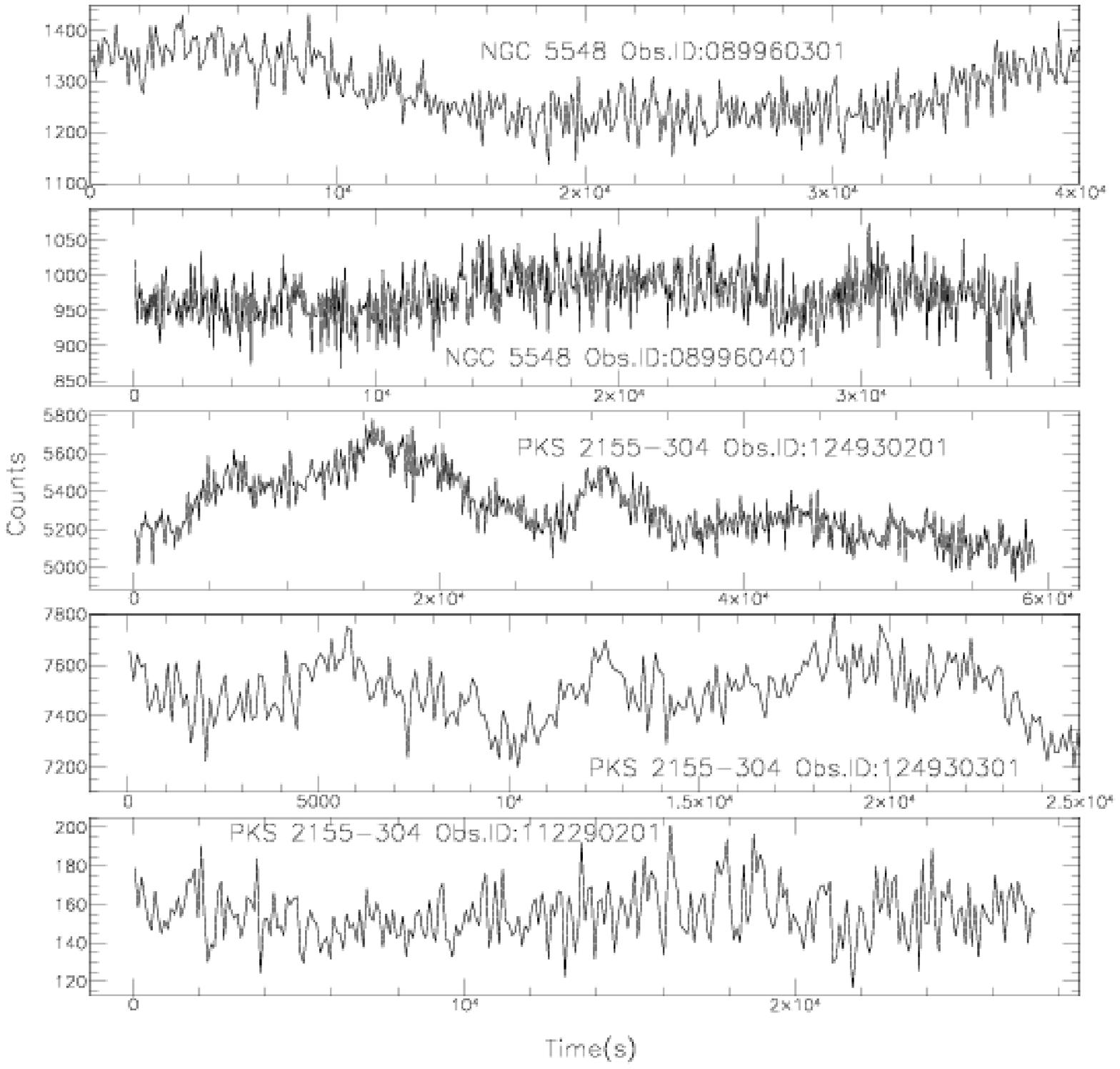}
%\plottwo{figure3.ps} {figure4b.ps}
\caption{Light curves for the 10 AGN in our sample.  All observations are in the energy range 0.75 to 10 keV.  Most observations are binned to 100s except for MKN 421, NGC 5548 (Obs ID: 089960401) and IRAS 13349+2438 which are binned to 50s.\label{lightcurves2}}
\end{figure}

\citet{fiore} report a QPO in NGC 4151 around 5.8 ks with $>99\%$ significance based on three EXOSAT observations.  \citet{vaub} reanalyzed these data sets and found that after fitting the red noise significance and Poisson photon noise components of the spectrum, the QPOs fall below the 95$\%$ threshold.  Our {\it XMM-Newton} observations show a $\sim$4.8 ks feature.  This appears in our 57 ks observation (Obs. ID: 112830201) with 96$\%$ red noise significance and in our 30 ks observation (Obs. ID: 112310101) with 99.4$\%$ red noise (see Fig.~\ref{gwpsall2}).  Even though this signal rises above the 99$\%$ red noise level in this observation, we discount it because it appears as part of a larger power structure in the GWPS and is not a well-defined peak.

For NGC 5548, \citet{pap93} claim a 500s QPO in five out of eight EXOSAT observations.  \citet{tag96} reanalyzed the same data and found that one observation had detector problems.  Also, in every case they found less than 95$\%$ significance by taking into account the uncertainties in modeling the spectrum.  In our {\it XMM-Newton} data of NGC 5548, we report a 500s feature,  but it has only a 93$\%$ red noise significance, and is seen in only one of our two observations (Obs. ID: 089960401, Fig.~\ref{gwpsall2}).

For MKN 766, \citet{boller} claim a $\sim$4200s QPO in a 30 ks {\it XMM-Newton} observation.  In our 128 ks observation, taken a year later, we see a signal at 4200s with 99.5$\%$ red noise significance, but it is dwarfed in the global wavelet power by a much stronger, wider broad peak (see Fig.~\ref{gwpsall2}), possibly due to a secular change in flux over the observation.

We do not detect any significant feature for MCG-6-30-15 \citep{lee}, MKN 421 or PKS 2155-304 \citep{osone}, which have previously reported QPOs.  To the best of our knowledge there are no published QPO claims for any of the other objects in our sample:  3C 273, IRAS 13349$+$2438, NGC 3516.

\citet{hal03} reported the discovery of a 2.08 day quasi-period in the NLS1 galaxy TonS180 with a 33 day observation taken with the  $Extreme$ $Ultraviolet$ $Explorer$ (EUVE).  \citet{vau} suggest  that this periodogram is oversampled and so the significance is overestimated.  Our wavelet  analysis of this data shows a $\sim~$2 d period in the global wavelet power spectrum which  rises above the 99$\%$ white noise level, but it has only 89.5$\%$ red noise significance (Fig.~\ref{gwpsall2}).  Our wavelet analysis finds an $\alpha$ of 0.8 for this observation and the structure function shows that red noise dominates, implying that this 2 d feature should be compared to red noise significance and so is not significant.

\section{Summary \& Conclusions}

We applied the wavelet analysis technique to {\it XMM-Newton} observations of 10 AGN and detected a candidate 3.3 ks period in 3C 273.  % with a significance of 99.979$\%$ red noise with $\alpha = 0.14$ and 99.99$\%$ white noise.  
The cross wavelet transform shows that the 3.3 ks signal is present throughout the entire observation.  
%We did not find any of the previously reported QPOs for the AGN in our sample at what we regard as a statistifcally significant level.

%The wavelet technique is superior to the Fourier transfrom in that the continuous wavelet transform tells us which frequencies exist in signal and when.  The cross-wavelet transform examines how two time series are correlated in time-frequency space. 
%Here, we use the cross-wavelet to compare a mock sinusoidal signal with a period of 3304 s with the 3C 273 lightcurve.  The Global Wavelet Power Spectrum,  which sums up the wavelet power spectra at all times, shows that the 3.3 ks detection exceeds the expected levels of white and red noise at the 99$\%$ significance level.  
%Monte Carlo simulations showed that the probability of a spurious detection with the wavelet technique in this case is 1$\%$.  %Also, as further evidence of the wavelet technique's robustness, it yields the same QPOs detected using Fourier methods in previously published X-ray observations of XTE J1550-564. 

%Although strong in the previously discussed observation of 3C 273, this periodicity is not detected in any of the other three observations of this object or in Fourier power spectra.  

If the 3.3 ks quasi-period in 3C 273 represents an orbital timescale originating near a last stable orbit, it implies a central black hole mass of at least $7.3\times 10^6$ M$_\sun$ which does not agree with reverberation mapping mass estimates.  \citet{kas00} estimate the mass of the black hole in 3C 273 at $2.35\times 10^8$ M$_\sun$ and \citet{paltani} find a mass of $6.59 \times10^{9}$ M$_\sun$.  This suggests that this X-ray quasi-period is not caused by dynamical motion in the inner accretion disk.  

We suggest that oscillations with modes of three or higher are occurring in the accretion disk of 3C 273, producing the detected 3.3 ks quasi-period.  \citet{perez} shows that for $m\geq3$ and maximum angular momentum one can obtain a mass for 3C 273 of $1.4 \times 10^8$ M$_\sun$, consistent with the lower mass estimate obtained from reverberation mapping.

\acknowledgments
We thank the anonymous referee for
constructive comments that helped improve the paper.  We thank Margo Aller, Robert Fender, Luis Ho, Jereon Homan, Jon Miller, and Simon Vaughan for useful discussions.
JNB would like to acknowledge support from NASA for these activities,
through the Long Term Space Astrophysics grant NAG5-10765.

\clearpage

\clearpage

\clearpage

\clearpage

\clearpage

\clearpage

\clearpage

\clearpage

\clearpage

\clearpage

\clearpage

\clearpage

\clearpage

\clearpage

\clearpage

\clearpage
\begin{deluxetable}{lcccc}
\tabletypesize{\scriptsize}
\tablewidth{0pt}
\tablenum{1}
\tablecolumns{5}
\tablecaption{Log of Observations\label{obslog}}
\tablehead{
\colhead{Object} & \colhead{Observation ID} & \colhead{Start Date} & \colhead{Length(ks)} & \colhead{$\langle$Counts$\rangle$}
}
\startdata
3C 373 & 126700301 & 2000-06-13 & 66 & 140 \\
3C 373 & 126700801 & 2000-06-17 & 60.6 & 120 \\
3C 373 & 136550101 & 2003-01-05 & 88.6 & 160 \\
3C 273 & 159960101 & 2003-07-07  & 58 & 230 \\
IRAS 13349+2438 & 096010101 & 2000-06-20 & 44.6 & 10 \\
M81 & 111800101 & 2001-04-22 & 130 & 20 \\
MCG-6-30-15 & 029740701 & 2001-08-02 & 127 & 70 \\
MCG-6-30-15 & 029740801 & 2001-08-04 & 125 & 120 \\
MKN 421 & 099280101 & 2000-05-25 & 32.5 & 740 \\
MKN 421 & 099280301 & 2000-11-13 & 46.6 & 1060 \\
MKN 766 & 109141301 & 2001-05-20 & 128.5 & 90 \\
NGC 3516 & 107460601 & 2001-04-10 & 129 & 20 \\
NGC 3516 & 107460701 & 2001-11-09 & 128 & 12 \\
NGC 4151 & 112310101 & 2000-12-21 & 30 & 20 \\
NGC 4151 & 112830201 & 2000-12-22 & 57 & 25 \\
NGC 5548 & 089960301 & 2001-07-09 & 93.4 & 75 \\
NGC 5548 & 089960401 & 2001-07-12 & 37 & 90 \\
PKS 2155-304 & 124930201 & 2000-05-31 & 59 & 280 \\
PKS 2155-304 & 124930301 & 2001-11-30 & 44.6 & 380 \\
\enddata
\tablecomments{Events are grouped in 5s bins.}
\end{deluxetable}

\clearpage
\begin{deluxetable}{lcccc}
\tabletypesize{\scriptsize}
\tablewidth{0pt}
\tablenum{2}
\tablecolumns{5}
\tablecaption{Structure Function Slopes and Timescales\label{sftbl}}
\tablehead{
\colhead{Object} & \colhead{Observation ID} & \colhead{Slopes} & \colhead{Transition Time (s)} & \colhead{Turnover Time (s)}
}
\startdata
3C 373 & 126700301 & 1.23 & $1.5\times10^4$ & -\\
3C 373 & 126700801 & 2.09 & $1.5\times10^4$ & -\\
3C 373 & 136550101 & - & - & -\\
3C 273 & 159960101 & 1.68 & $2\times10^4$& - \\
IRAS 13349+2438 & 096010101 & 1.68 & 1000 & - \\
M81 & 111800101 & 0.95 & 7000 & -\\
MCG-6-30-15 & 029740701 & 0.82 & - & $10^4$ \\
MCG-6-30-15 & 029740801 & 1.11 & 200 & - \\
MKN 421 & 099280101 & 1.25 & 400 & - \\
MKN 421 & 099280301 & 1.13 & 350 & 7000 \\
MKN 766 & 109141301 & 0.67 & 100 & $3\times10^4$ \\
NGC 3516 & 107460601 & 1.18,1.39 & 2000 & - \\
NGC 3516 & 107460701 & 1.86 & $2\times10^4$ &- \\
NGC 4151 & 112310101 & - & - & - \\
NGC 4151 & 112830201 & 1.19 & 7000 & - \\
NGC 5548 & 089960301 & 0.99,1.64 & 1200 & - \\
NGC 5548 & 089960401 & 0.097,2.75 & 1000 & - \\
PKS 2155-304 & 124930201 & .99,.59,.45,1.69 & 2000 & - \\
PKS 2155-304 & 124930301 & 1.59 & 1100 & - \\
\enddata
\tablecomments{Transition time corresponds to the time at which the SF curve changes from plateau to power-law.  Turnover time is the time at which the power-law portion of the SF curve changes to a plateau. }
\end{deluxetable}

\end{document}